# Predicting CaO-(MgO)-Al₂O₃-SiO₂ glass reactivity in alkaline environments from force field molecular dynamics simulations

Kai Gong[1,2], Claire E. White[1,*]

[1]Department of Civil and Environmental Engineering and the Andlinger Center for Energy and the Environment, Princeton University, Princeton, New Jersey 08544, United States

[2]Department of Materials Science and Engineering, Massachusetts Institute of Technology, Cambridge, MA 02139, United States (current address)

* Corresponding author: Phone: +1 609 258 6263, Fax: +1 609 258 2799, Email: whitece@princeton.edu

Postal address: Department of Civil and Environmental Engineering, Princeton University, Princeton NJ 08544, USA

Keywords: Molecular Dynamics Simulations; Amorphous Aluminosilicate; Structural Descriptors; Glass Reactivity; X-ray Scattering; Neutron Scattering; Granulated Blast-Furnace Slag

## Abstract

In this investigation, force field-based molecular dynamics (MD) simulations have been employed to generate detailed structural representations for a range of amorphous quaternary CaO-MgO-Al₂O₃-SiO₂ (CMAS) and ternary CaO-Al₂O₃-SiO₂ (CAS) glasses. Comparison of the simulation results with select experimental X-ray and neutron total scattering and literature data reveals that the MD-generated structures have captured the key structural features of these CMAS and CAS glasses. Based on the MD-generated structural representations, we have developed two structural descriptors, specifically (i) average metal oxide dissociation energy (AMODE) and (ii) average self-diffusion coefficient (ASDC) of all the atoms at melting. Both structural descriptors are seen to more accurately predict the relative glass reactivity than the commonly used degree of depolymerization parameter, especially for the eight synthetic CAS glasses that span a wide compositional range. Hence these descriptors hold great promise for predicting CMAS and CAS glass reactivity in alkaline environments from compositional information.

# 1 Introduction

Amorphous aluminosilicates are of significant interest to many technologically important fields and applications, including geology, glass science, metallurgical process, nuclear waste encapsulation and sustainable cement production. In particular, various amorphous aluminosilicates have been used as precursor sources to synthesize the so-called alkali-activated materials (AAMs), an important class of low-$CO_2$ cement-based binder [1, 2]. Alternatively, amorphous aluminosilicates are commonly used as supplementary cementitious materials (SCMs) in blended cement to partially replace Portland cement [3, 4] and hence lower the carbon footprint of the cement industry (currently responsible for 8-9% of global anthropogenic $CO_2$ emissions) [5]. Most of the commonly used amorphous aluminosilicates for the above two applications are industrial by-products (e.g., ground granulated blast-furnace slag (GGBS) and coal-derived fly ash), although other sources of amorphous aluminosilicates are being actively explored, especially calcined clays, which are attractive due to the extremely large clay reserves [6].

The chemical composition, minerology and particle size of these precursor materials and SCMs can vary considerably depending on their type, source location and processing parameters. Even for GGBSs, which have relatively small chemical variability compared to fly ash, their main oxide components do vary, consisting of CaO (30– 50 wt.%), $SiO_2$ (28–38 wt.%), $Al_2O_3$ (8–24 wt.%) and MgO (1–18 wt.%) along with the presence of other trace elements (e.g., S, Ti, Na, K, Mn and Fe) as well as crystalline impurities (e.g., merwinite, gehlenite, åkermanite, calcite and quartz) [4, 7-11]. These inherent variabilities can have a dramatic impact on precursor/SCM reactivity in both AAM and blended Portland cement systems, as well as the resulting pore structure and engineering properties of the final cementitious product [1, 3, 7, 8, 11-13]. The impact of Ca content is particularly profound. First, Ca is a known network modifier and tends to increase the framework disorder and the degree of depolymerization (and hence the reactivity) of the glassy aluminosilicate precursor [14]. Recent investigations on synthetic glasses have shown that Ca-rich aluminosilicate glass exhibits a significantly higher reactivity than Si-rich counterparts [15-18]. This is a major reason why Ca-rich GGBS can achieve a higher replacement ratio in blended Portland cements [4] and be activated using (i) a much lower alkali content in sodium hydroxide or sodium silicate activators or (ii) weak activators (e.g., $Na_2CO_3$ and $Na_2SO_4$) for AAM applications, as compared with low-Ca precursors (e.g., class F fly ash and metakaolin) [1].

Calcium is also important when it comes to the atomic structure, transport properties and long-term durability of the precipitated binder gel in AAMs [1, 19]. At low Ca content (i.e., class F fly ash and metakaolin), the alkali activation reaction results in a three-dimensional alkali-alumino-silicate-hydrate gel (N-A-S-H gel if sodium is the alkali) with predominately $Q^4$ silicate units ($Q^n$ denotes $n$ bridging oxygens) [1, 2]. In contrast, for AAMs based on a Ca-rich precursor (e.g., GGBS and class C fly ash), the resulting binder gel is an alkali-containing calcium-alumino-silicate-hydrate gel (C-(N)-A-S-H gel if sodium is the alkali) dominated by a depolymerized chain-like silicate structure [2, 20-23], similar to the calcium-silicate-hydrate (C-S-H) and calcium-alumino-silicate-hydrate (C-A-S-H) gel in Portland cement and blended Portland cement systems containing aluminum. This difference in the binder gel is linked with noticeable differences in pore structures [24-27], transport properties [25, 26, 28, 29] and chemical stability [26, 28].

The impact of alumina content on the reactivity of amorphous aluminosilicates and engineering properties of the resulting AAM and blended Portland cements has also been investigated [11, 15]. In a 2014 review article, Provis and Bernal suggested that high Al content is beneficial to the strength development of fly ash-based AAMs, similar to the impact of the network modifier (e.g., alkali and alkali earth metal) content [2]. A recent investigation on synthetic calcium aluminosilicate (CAS) glasses showed that an increasing Al content (at fixed Ca content) leads to a higher extent of reaction in a blended mixture of portlandite, limestone and sodium hydroxide [15]. In contrast, an earlier investigation on the reaction kinetics of NaOH- and $Na_2SiO_3$-activated GGBSs showed that GGBS with a higher $Al_2O_3$ content leads to slower reaction kinetics and lower compressive strength during the early stages of reaction [11].

Magnesium has also been investigated, specifically regarding its impact on the reactivity of amorphous aluminosilicates. Ben Haha *et al.* examined three GGBS sources with different MgO content (8-13 wt. %) and found that a higher MgO content accelerates the early stage of reaction for alkali-activated GGBS (more apparent when $Na_2SiO_3$ activator was used) and increases compressive strength [12]. This is consistent with another investigation on $Na_2CO_3$-activated GGBS [8], where GGBS with a higher MgO content was seen to exhibit much faster reaction kinetics as evaluated using isothermal conduction calorimetry (ICC). The beneficial impact of MgO on compressive strength observed by Ben Haha *et al.* was also in agreement with an earlier investigation by Douglas *et al.* [30], which showed that the 28-day compressive strength of silicate-activated GGBS triples when the MgO content of GGBS increases from 9 to 18 wt. %. In contrast, another investigation on $Na_2SiO_3$-activated GGBS showed that GGBS with a lower MgO content reacts faster during the early stages of reaction [9]. As suggested in ref. [9], this inconsistency associated with the impact of MgO content on the reaction kinetics is related to the differences in the $Al_2O_3$ content of the GGBSs. Dissolution experiments on synthetic quaternary $CaO$-$MgO$-$Al_2O_3$-$SiO_2$ (CMAS) glasses showed that increasing the Mg/Ca ratio whilst maintaining a relatively fixed Si and Al content leads to a higher dissolution rate in aqueous solutions with pH of up to ~12 (especially in acid conditions) [16]. This observation for CMAS glass is inconsistent with silicate mineral dissolution experiments which generally show that the dissolution rate of dimagnesium silicate is several orders in magnitude lower than that of dicalcium silicate [31, 32]. The different impact Ca and Mg atoms have on the reactivity of silicate glasses and minerals could be associated with the formation of highly reactive free oxygen (FO) sites in CMAS glasses (not present in common silicate minerals), which seem to be promoted by Mg atoms [33, 34].

Despite it being clear that the composition of an amorphous calcium/magnesium aluminosilicate has a significant impact on (i) its reactivity as an SCM in blended Portland cements or a precursor material for AAMs, and (ii) the final properties of the cementitious product, there have only been a limited number of investigations on the composition-structure-properties relationship for these amorphous aluminosilicates. Many investigations (including refs. [8, 9, 11, 12] discussed above) focus on describing the composition-properties relationship using individual oxide components or empirical reactivity index (e.g., $(CaO+MgO)/SiO_2$ from European Standard for slag cement) [35, 36]. Several recent investigations [15, 17, 37] have used the degree of depolymerization (i.e., the number of non-bridging oxygen (NBO) per network former T (NBO/T), where T = Si and Al atoms in IV-fold coordination) of the glassy phase as a structural descriptor, which is commonly used in the glass community and can be estimated from chemical composition based on classical glass theory [38]. These investigations [15, 17] have generally showed a positive correlation between the degree of depolymerization (or NBO/T) of the glass and its reactivity in an alkaline environment. However, several investigations have suggested that NBO/T is not always a reliable indicator of glass reactivity [15, 39]. For instance, in ref. [15], a decrease of reactivity with increasing NBO/T has been observed for several synthetic CAS glass compositions relevant to fly ash. NBO/T has also been used to describe mineral dissolution, where a general positive trend (between NBO/T and dissolution rate) has been observed [31]. However, it has also been shown for alkali earth metal silicate minerals that the dissolution rate can vary several orders of magnitude for minerals with the same level of NBO/T [31, 32].

The inability for individual oxide components (e.g., $Al_2O_3$ and MgO) or the commonly used NBO/T parameter to accurately predict GGBS or C(M)AS glass reactivity in AAMs and blended cements shows that there is a need to develop more reliable structural descriptors to connect the composition of these amorphous aluminosilicates to their reactivity and associated final properties of the cement-based materials. Although it is challenging to obtain structural information on amorphous aluminosilicates, several experimental techniques have been shown to be extremely valuable, including nuclear magnetic resonance (NMR) [15, 40, 41], and X-ray and neutron total scattering [7, 33, 42-44]. On the other hand, atomistic modeling techniques, including force field-based molecular dynamics (MD) simulations [34, 43, 45, 46] and quantum mechanics-based density functional theory (DFT) calculations [47, 48], have been successfully used to generate detailed and realistic structural representations for aluminosilicate glasses, including when combined with X-ray and neutron scattering experiments [33]. Furthermore, MD simulations have been recently employed in the glass community to derive structural information that connects glass composition and molecular features to glass properties, including Young's modulus, density, viscosity, glass transition temperature, and leaching and chemical durability [49]. However, similar MD investigations linking composition-structure-properties for quaternary CMAS and ternary CAS glasses that are representative of SCMs and AAM precursors are rare.

In this investigation, force field MD simulations have been employed to generate detailed structural representations for 18 CMAS and CAS glasses with a wide range of compositions related to GGBSs/glasses that were previously studied in four high-quality experimental investigations [8, 11, 12, 15]. Detailed analysis of MD-derived structures has been carried out to determine their structural attributes (including the nearest interatomic distances, coordination numbers and the degree of depolymerization), which were subsequently compared with (i) our X-ray and neutron total scattering data collected on select GGBS compositions, (ii) literature data, and/or (iii) theoretical estimation, to ensure that the structural representations generated were reasonable. Based on the MD simulation results, two structural descriptors have been derived, i.e., (i) the average metal oxide dissociation energy (AMODE) and (ii) average self-diffusion coefficient (ASDC) of all the atoms at melting, and their performance in predicting the reactivity data from the experimental investigation has been evaluated, in comparison with the commonly used NBO/T parameter (i.e., the degree of depolymerization) also derived from MD simulations. This investigation serves as a crucial step forward in establishing the important composition-structure-reactivity relationship for amorphous aluminosilicates in alkaline environments, relevant to blended Portland cements and AAMs.

## 2    Methodology

### 2.1    Glass compositions

We selected ten GGBSs composed of predominantly CMAS glassy phases and eight synthetic CAS glasses with a range of chemical compositions from four separate high-quality investigations [8, 11, 12, 15], where each investigation experimentally investigated the impact of glass composition on the reactivity in alkaline conditions. The chemical compositions and physical properties of the CMAS and CAS glasses from these investigations are summarized in Table 1. All the GGBSs (Group A-C in Table 1) are predominately amorphous (as evidenced by the X-ray diffraction (XRD) data in each investigation) with ~94-96 wt. % of CMAS glass and 3-5 wt. % of other minor oxide phases (e.g., $SO_3$, $K_2O$, $Na_2O$, $TiO_2$ and $Mn_2O_3$) [8, 11, 12]. All the GGBSs contain ~34-43 wt. % CaO, ~31-42 wt. % $SiO_2$, ~7-17 wt.% $Al_2O_3$ and ~1-14% wt.% MgO. Although the compositional variation is relatively small, especially for the two major oxide components (CaO and SiO$_2$), significant differences in reactivity have been observed in the experimental investigations, especially in ref. [8], where the impact of MgO content was studied (Group A in Table 1). The two other investigations focused on the impact of MgO (Group B in Table 1) [12] and Al$_2$O$_3$ content (Group C in Table 1) [11] in the GGBSs on their reactivity during alkaline activation, however, the quantities of the different oxide components in each group (A-C) appear to be interconnected as illustrated in Figure 1. It is clear from Figure 1 that CaO, MgO and Al$_2$O$_3$ content are strongly correlated with the SiO$_2$ content, especially for the GGBSs in Group B and C ($R^2$ values close to 1.00 for linear fits). Similarly, strong correlations are observed between the CaO content and MgO and Al$_2$O$_3$ content for these GGBSs, with the results shown in Figure S1 of the Supplementary Material. In fact, in our previous investigation on seven GGBSs from different origins, we also observed that the main compositions of these GGBSs are interconnected [7]. Therefore, the different levels of reactivity in each group of GGBSs, as observed in refs. [8, 11, 12], should not be simply attributed to their compositional difference in one oxide component (e.g., MgO or Al$_2$O$_3$). For a more accurate description of composition-reactivity relationship for these GGBSs it is necessary to first obtain detailed atomic structural information, as has been carried out in this study.

Table 1. The chemical composition of the main oxides (in weight percentage), particle surface area and density of the different GGBSs and synthetic glasses from refs. [8, 11, 12, 15]. Note that the uncertainty associated with surface area data was only reported for the GGBSs in Group A.

| ID # | CaO | MgO | SiO$_2$ | Al$_2$O$_3$ | Surface area (cm$^2$/g) | Notes and sources |
|---|---|---|---|---|---|---|
| A1_1Mg | 42.9 | 1.2 | 31.6 | 14.6 | $4012 \pm 49$ | Investigated the impact of GGBS Mg |
| A2_5Mg | 42.3 | 5.2 | 32.3 | 13.3 | $4435 \pm 109$ | content on its reactivity during |
| A3_7Mg | 41.3 | 6.5 | 36.0 | 11.3 | $5056 \pm 22$ | Na$_2$CO$_3$ activation [8] |
| A4_14Mg | 33.9 | 14.3 | 37.4 | 9.0 | $4794 \pm 44$ | |
| B1_8Mg | 35.8 | 7.7 | 38.2 | 12.0 | 4990 | Investigated the impact of GGBS Mg |
| B2_11Mg | 34.6 | 10.5 | 37.1 | 11.5 | 5070 | content on its reactivity during |
| B3_13Mg | 33.4 | 13.2 | 36.4 | 11.3 | 5010 | NaOH and Na$_2$SiO$_3$ activation [12] |
| C1_7Al | 39.1 | 7.2 | 41.6 | 7.0 | 5021 | Investigated the impact of GGBS Al |
| C2_14Al | 36.0 | 6.6 | 38.2 | 14.1 | 4963 | content on its reactivity during NaOH |
| C3_17Al | 35.0 | 6.4 | 37.2 | 16.7 | 4985 | and $Na_2SiO_3$ activation [11] |
| D1 | 4.7 | 0.0 | 78.5 | 16.8 | 4720 | Investigated the impact of CAS |
| D2 | 4.7 | 0.0 | 69.1 | 26.3 | 4810 | synthetic glass composition on its |
| D3 | 4.3 | 0.0 | 60.6 | 35.1 | 4800 | reactivity in a mixture of $Ca(OH)_2$, |
| D4 | 13.9 | 0.0 | 59.4 | 26.7 | 4550 | NaOH, and limestone [15] |
| D5 | 21.4 | 0.0 | 62.0 | 16.6 | 4630 | |
| D6 | 24.1 | 0.0 | 49.8 | 26.1 | 4680 | |
| D7 | 24.0 | 0.0 | 39.7 | 36.3 | 4220 | |
| D8 | 49.9 | 0.0 | 34.8 | 15.3 | 3960 | |

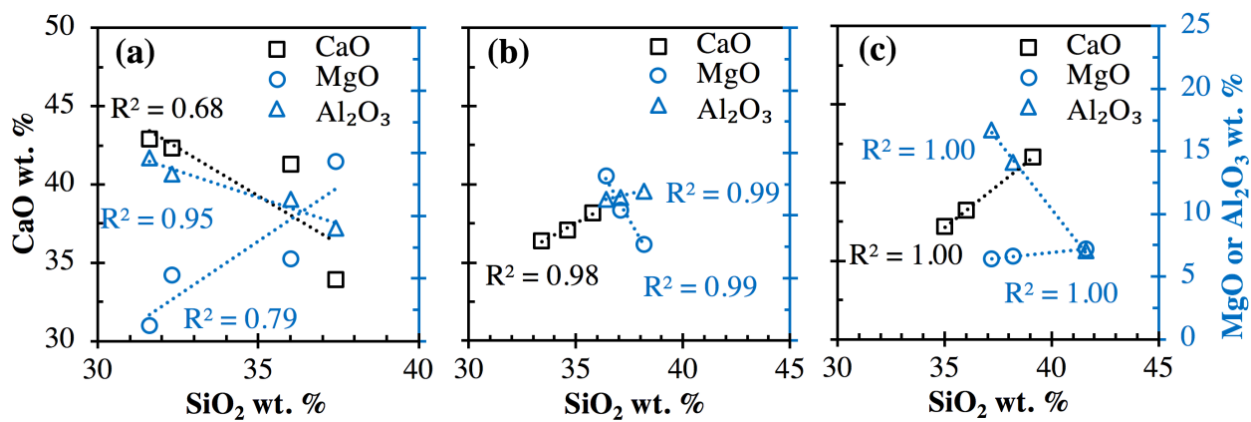

Figure 1. Correlation between $SiO_2$ content and CaO, MgO and $Al_2O_3$ content for the GGBS compositions in (a) Group A, (b) Group B and (c) Group C (see Table 1 for the compositions). $R^2$ values for linear fits are given in the figure.

The eight CAS compositions in Group D are for synthetic glasses selected from another investigation [15] that span a wider compositional range (~4-50 wt. % CaO, ~35-79 wt. % $SiO_2$ and ~15-36 wt. % $Al_2O_3$) than the CMAS glasses in Group A-C. Simple analysis shows that the correlation between CaO content and $SiO_2$ and $Al_2O_3$ content in this group ($R^2$ values of 0.78 and 0.13, respectively; see Figure S2 in Supplementary Material for details) is much weaker than those in Group A-C. In the original investigation, these CAS glass compositions were designed to uncover the impact of $Al_2O_3$ at two CaO levels, i.e., D1-3 and D5-7 glasses with targeted CaO content of 5.0 and 25.0 wt. %, respectively (the values in Table 1 are the experimentally obtained composition data). The former three compositions (i.e., D1-3) are relevant to Si-rich fly ash (e.g., class F) while the latter three (i.e., D5-7) are relevant to Ca-rich fly ash (e.g., class C). The D8 glass was designed to represent a GGBS composition without MgO. D2, D4 and D6 compositions were designed to investigate the impact of Ca/Si ratio at fixed $Al_2O_3$ content (i.e., ~26 wt. %). The XRD patterns in ref. [15] show that these synthetic CAS glasses are predominantly amorphous.

Table 1 also includes specific surface area data for all the glasses, which exhibit ~2-20% difference within each group, although efforts have been made in each experimental investigation to ensure similar particle size distributions [8, 11, 12, 15]. This difference in specific surface area needs to be considered when evaluating glass reactivity; for example, a recent investigation has shown that the reactivity of GGBS glasses in alkaline environments (based on ICC measurements) increases almost linearly with specific surface area ($R^2$ values of 0.97-1.00 for linear fits) [36]. Hence, the reactivity data extracted from refs. [8, 11, 12, 15] have been normalized by the particle surface area of each glass prior to evaluation of the relative glass reactivity within each group.

## 2.2   Computational details

Force field MD simulations have been used in this investigation to generate amorphous structural representations for all the CMAS and CAS glass compositions shown in Table 1, using simulation cells consisting of ~2000 atoms. The force field parameters developed by Guillot for crystals and melts of the $CaO-MgO-Al_2O_3-SiO_2$ system were used for all the simulations [50]. The simulations were performed using the ATK-Forcefield module in the Virtual NanoLab (VNL) software package [51, 52] following the commonly used melt-and-quench approach, similar to the MD simulation section in our previous work on a CMAS glass [33].

Specifically, we started with random structures in cubic cells with the same CMAS or CAS compositions as the experimental data in Table 1 (elemental compositions are shown in Table 2). For each structure, the density of the unit cell was initially set at a value estimated for CMAS glass at a temperature of 5000 K. The value was estimated using a similar method adopted in our previous investigation [33], and detailed calculations are given in Section 2 of the Supplementary Material. The structure was firstly equilibrated at 5000 K for 1 ns to ensure the loss of the memory of the initial configuration. It was then quenched from 5000 to 2000 K over 3 ns followed by equilibration at 2000 K for 1 ns, before being further quenched from 2000 to 300 K over 3 ns and equilibrated at 300 K for 1 ns. The canonical *NVT* ensemble with the Nosé Hoover thermostat and a time step of 1 fs were used for all the MD simulation steps mentioned above, while the density of the unit cell (i.e., volume) was adjusted to numerically estimated or experimental values (as shown in Table 2) at the start of each equilibration step. For the eight synthetic CAS glasses (Group D), experimental density values at room temperature are available in ref. [15] and hence were used for the equilibration step at 300 K (see Table 2). However, for the CMAS glasses in Group A-C, room temperature density values are only available for some of the corresponding GGBSs [8, 11, 12]. Hence, the densities of the CMAS glasses in Group A-C at 300 K and all the higher temperature densities for all glasses (Group A-D) were numerically estimated using the method presented in Section 2 of the Supplementary Material. In summary, the estimated densities of the final structures at 300 K for the CMAS glasses (~2.81-2.88 $g/cm^3$; shown in Table 2) are within ~3% of the experimental values of GGBSs with similar $CaO-MgO-Al_2O_3-SiO_2$ compositions (~2.87-2.94 $g/cm^3$) [11, 12]. Two configurations during the last 500 ps of the MD equilibration step at 300 K (separated by 500 ps) were extracted, and the whole process was repeated three times to generate six structural representations for each of the eighteen CMAS and CAS compositions given in Table 2. These structural representations were further analyzed to obtain the proportion of different oxygen species (including NBO and FO). For three GGBSs in Group A with available experimental PDF data (experimental details outlined in next section), the corresponding structural representations were used to generate simulated PDFs for comparison with the experimental data. Note that all the GGBSs also contain trace amounts of minor oxides (< 3-5 wt. %), which were not included in the simulation due to their relatively small quantities, as explained in more detail in our previous investigation [33].

Table 2. The number of atoms in each simulation box (corresponding to the oxide composition of each GGBS or synthetic glass shown in Table 1) along with the numerically estimated or experimentally-determined cell density (labeled with $^*$) used at each equilibration temperature during the MD simulations. The numerical calculation method for the density values at different temperatures is given in Section 2 of the Supplementary Material. The theoretical degree of depolymerization (i.e., NBO/T) has been calculated based on simple stoichiometric considerations [38], as explained in detail in Section 3 of the Supplementary Material.

| Glass ID # | Number of atoms in the simulation box | | | | | | Density (g/cm$^3$) at given temperature (K) | | | Theoretical NBO/T |
|---|---|---|---|---|---|---|---|---|---|---|
| | Ca | Mg | Si | Al | O | Total | 300 | 2000 | 5000 | |
| A1_1Mg | 394 | 15 | 271 | 148 | 1173 | 2001 | 2.86 | 2.69 | 2.39 | 2.20 |
| A2_5Mg | 375 | 65 | 267 | 130 | 1169 | 2006 | 2.87 | 2.70 | 2.40 | 1.89 |
| A3_7Mg | 355 | 78 | 290 | 106 | 1172 | 2001 | 2.86 | 2.69 | 2.39 | 1.92 |
| A4_14Mg | 286 | 169 | 294 | 82 | 1166 | 1997 | 2.85 | 2.68 | 2.38 | 1.60 |
| B1_8Mg | 307 | 93 | 306 | 114 | 1183 | 2003 | 2.82 | 2.65 | 2.35 | 1.63 |
| B2_11Mg | 296 | 126 | 296 | 108 | 1176 | 2002 | 2.83 | 2.66 | 2.36 | 1.82 |
| B3_13Mg | 282 | 156 | 287 | 104 | 1168 | 1997 | 2.84 | 2.67 | 2.37 | 1.97 |
| C1_7Al | 334 | 86 | 332 | 66 | 1183 | 2001 | 2.81 | 2.64 | 2.34 | 1.94 |
| C2_14Al | 305 | 78 | 302 | 132 | 1185 | 2002 | 2.82 | 2.65 | 2.35 | 1.46 |
| C3_17Al | 294 | 75 | 292 | 154 | 1184 | 1999 | 2.82 | 2.65 | 2.35 | 1.31 |
| D1 | 34 | 0 | 532 | 134 | 1299 | 1999 | 2.49$^*$ | 2.31 | 2.00 | -0.10[†] |
| D2 | 34 | 0 | 469 | 210 | 1287 | 2000 | 2.59$^*$ | 2.35 | 2.05 | -0.2[†] |
| D3 | 31 | 0 | 412 | 280 | 1275 | 1998 | 2.56$^*$ | 2.39 | 2.09 | -0.32[†] |
| D4 | 104 | 0 | 414 | 220 | 1262 | 2000 | 2.61$^*$ | 2.43 | 2.13 | -0.02 |
| D5 | 163 | 0 | 442 | 140 | 1257 | 2002 | 2.72$^*$ | 2.44 | 2.14 | 0.32 |
| D6 | 186 | 0 | 358 | 222 | 1235 | 2001 | 2.85$^*$ | 2.51 | 2.21 | 0.26 |
| D7 | 185 | 0 | 286 | 308 | 1219 | 1998 | 2.85$^*$ | 2.57 | 2.27 | 0.10 |
| D8 | 417 | 0 | 272 | 140 | 1171 | 2000 | 2.93$^*$ | 2.70 | 2.34 | 1.68 |

$^*$ Experimental density values from ref. [15].

¹Peraluminous region with no NBO in theory. The negative values indicate that there are insufficient Ca cations to charge balance all the Al atoms, assuming all Al atoms are in IV-fold coordination. A more negative value indicates a greater Ca cation deficiency.

## 2.3    Experimental details

X-ray and neutron total scattering data have been collected on several GGBS compositions in Group A, specifically A1_1Mg, Al_5Mg and Al_14Mg in Table 1. The data for Al_5Mg GGBS have already been presented in our previous study [33]. The X-ray total scattering data were collected at room temperature using the 11-ID-B beam line [53] at the Advanced Photon Source, Argonne National Laboratory, while neutron total scattering data were collected at the Lujan Neutron Scattering Center, Los Alamos National Laboratory, using the NPDF instrument [54]. The data collection and processing procedures for the total scattering data are similar to those adopted in our previous investigations [7, 33, 55]. Briefly, the pair distribution function (PDF), *G(r)*, was calculated by taking a sine Fourier transform of the measured total scattering function, *S(Q)*, where $Q$ is the momentum transfer, as outlined by Egami and Billinge [56]. More details on the calculation of the PDF are given in Section 4 of the Supplementary Material. The X-ray PDF data were generated following a standard data reduction procedure using PDFgetX3 [57], with a $Q_{max}$ of 20 Å$^{-1}$. X-ray instrument parameters ($Q_{broad}$ = 0.016 Å$^{-1}$ and $Q_{damp}$ = 0.035 Å$^{-1}$) were obtained by using the calibration material (nickel, Sigma-Aldrich) and the refinement program PDFgui [58]. The PDFgetN software [59] and a $Q_{max}$ of 20 Å$^{-1}$ were used for the generation of the neutron PDF, where a background subtraction to remove incoherent scattering has been carried out [60]. The neutron instrument parameters ($Q_{broad}$ = 0.00201 Å$^{-1}$ and $Q_{damp}$ = 0.00623 Å$^{-1}$) were obtained using a silicon calibration material and the refinement program PDFgui [58]. These instrument parameters were used in PDFgui to compute the simulated PDFs based on the MD-generated structural representations for comparison with the corresponding experimental X-ray and neutron PDF data.

## 3 Results and Discussion

### 3.1 Comparison of structural representations with experimental data

The feasibility of the atomic structural representations obtained using the simulated melt-quench method with force field MD simulations (as outlined in Section 2.2) is highly dependent on the accuracy of the adopted force field. Although the force field used in this study was parameterized to cover silicate crystals and melts including the CaO-MgO-Al$_2$O$_3$-SiO$_2$ system [50], it is necessary to assess whether the obtained structural representations can reasonably capture the structural features in the experimental data, given that there are obvious discrepancies between simulation and experimental synthesis conditions (i.e., quenching rates, as will be briefly discussed in this section).

### 3.1.1 CMAS glasses (Group A-C)

The ten CMAS glasses in Group A-C (shown in Tables 1 and 2) represent the level of chemical variation of the main glassy phase found in amorphous GGBS, which generally resides in the highly percalcic region ((CaO+MgO)/Al$_2$O$_3$ >1). In this region, there is a high proportion of excess modifier cations (i.e., Ca and Mg cations) available to create NBO species (defined as an O atom bonded with only one network former, Si or Al atom, within its first coordination shell) beyond those required to charge-balance the negative charges associated with 4-fold alumina (i.e., [Al(O$_{1/2}$)$_4$]$^{-1}$). Due to the high modifier content in Group A-C, these CMAS glasses have a relatively high extent of depolymerization (NBO/T of ~ 1.6 to 2.2 shown in Table 2), estimated from simple stoichiometric considerations [38] (Section 3 of the Supplementary Material for more details) that include: (i) both Si and Al atoms are network formers in IV-fold coordination, and (ii) each excess alkaline earth cation creates two NBOs (as each NBO receive one electron from the network former and hence has a charge of -1 in theory). Figure 2a shows a typical atomic structural representation for a CMAS glass (i.e., A2_5Mg CMAS composition in Table 1) which is clearly a highly disordered aluminosilicate network structure. Analysis of this structure gives an NBO/T value of ~1.76 (stdev ≈ 0.007, based on the six structural representations for this composition), which is reasonably close to the theoretical estimation from simple stoichiometric considerations (i.e., 1.89, as shown in Table 2) [38] and that obtained from DFT-optimized structures for the same CMAS composition in our previous study (i.e., 1.80) [33].

The simulated X-ray and neutron PDFs obtained using the structural representation in Figure 2a are compared with the corresponding experimental X-ray and neutron PDF data in Figure 2b and 2c, respectively, where the experimental data were collected on an amorphous GGBS with the same CMAS composition. It is clear from this figure that the structure generated using MD simulations can capture reasonably well (i) the amorphous nature of the CMAS glass (as evidenced by the absence of noticeable ordering above ~10 Å), and (ii) the short-range (< ~3 Å) and mid-range (~3-10 Å) ordering. The level of agreement achieved with the X-ray PDF data (as indicated by the $R_w$ value; 0.46) is not as good as that achieved with DFT calculations in our previous study on the same CMAS glass composition ($R_w$ of 0.35), where a smaller $R_w$ value implies better agreement ($R_w$ value defined in PDFgui software [58]; detailed calculation of $R_w$ value is given in Section 4 of the Supplementary Material). On the other hand, the MD-generated structure gives slightly better agreement with the neutron PDF data ($R_w = 0.31$) than the DFT-optimized structure ($R_w = 0.35$) due to the slight over-estimation of the nearest O-O interatomic distance from the PBE exchange-correlation functional used in the DFT calculations [33].

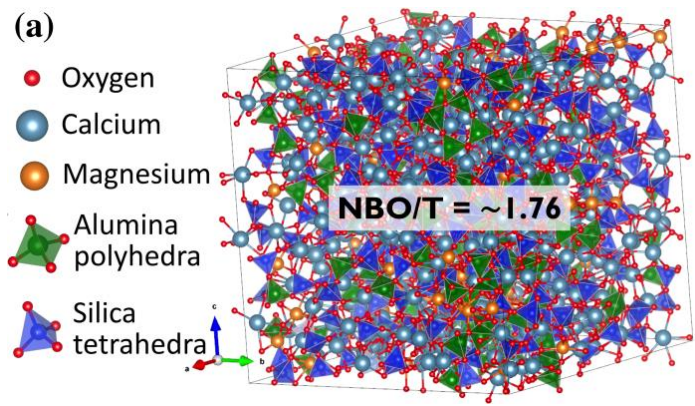

**(a)**
- Oxygen
- Calcium
- Magnesium
- Alumina polyhedra
- Silica tetrahedra

NBO/T = ~1.76

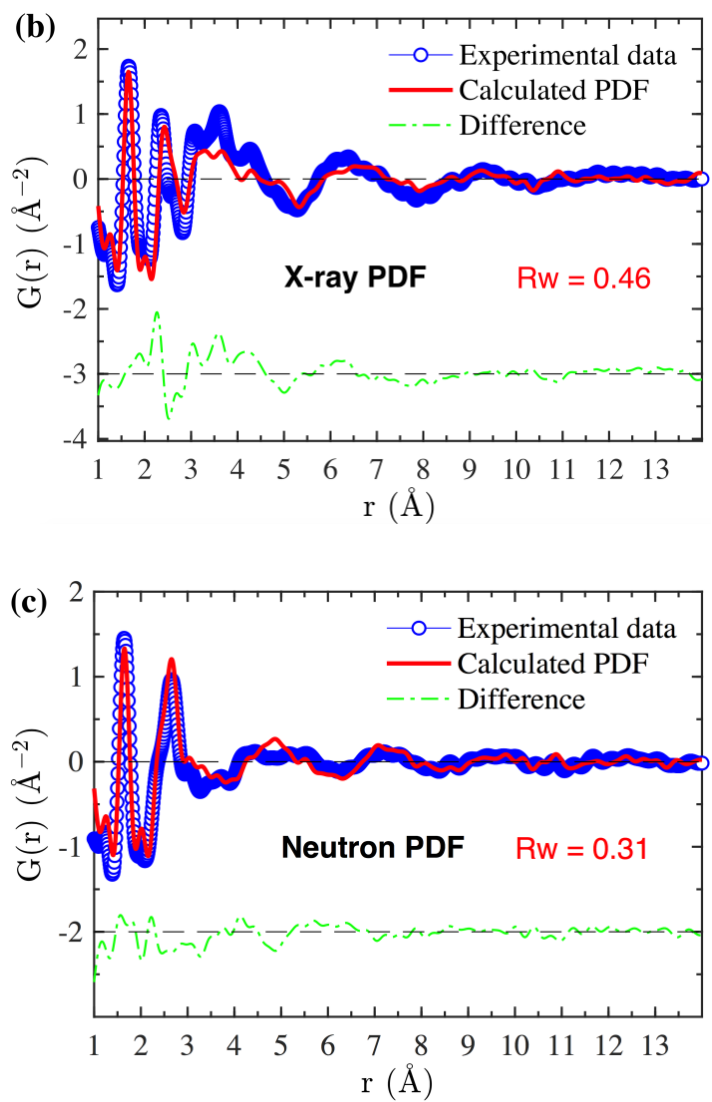

Figure 2. (a) A typical atomic structural representation obtained for a CMAS glass (i.e., A2_5Mg CMAS in Table 2), and the comparison between the simulated PDFs from a MD-generated atomic structural representation (shown in (a)) and the experimental (b) X-ray and (c) neutron PDF data of the corresponding GGBS with the same CMAS composition. NBO/T in (a) is the average number of non-bridging oxygen (NBO) species per network former T (T = Si and Al) calculated from the six structural representations of A2_5Mg CMAS. The level of agreement, as gauged by the $R_w$ values (refer to Section 4 in Supplementary Material for the calculation of $R_w$) is shown in (b) and (c).

We used force field MD simulations here (as opposed to more accurate DFT calculations) because the MD simulations allow for exploration of larger structures at a much lower computational cost while still capturing the key features of the CMAS glass structure (as evidenced in Figure 2b-c). We achieved similar levels of agreement with X-ray and neutron PDF data for two other CMAS glasses (i.e., A1_1Mg and A4_14Mg in Table 1), with the corresponding $R_w$ values summarized in Figure 3. Direct comparison of the simulated and experimental PDF data (X-ray and neutron), similar to Figure 2b-c, is given in Figure S3 of the Supplementary Material. The $R_w$ values for all the three samples are in the range of 0.44-0.47 and 0.30-0.32 for X-ray and neutron PDF data (Figure 3), respectively, indicating that the level of agreement between the experimental and simulated PDFs is similar to those shown in Figure 2b-c. The level of agreement is also comparable with several previous investigations on the modeling of the atomic structure of amorphous GGBS (0.35 for X-ray and 0.36 for neutron data) [33], iron-rich slag (0.38 for X-ray and 0.31 for neutron data) [43], magnesium carbonate ($R_w \approx 0.48$) [61] and metakaolin ($R_w \approx 0.77$) [42].

Nevertheless, similar to previous modeling investigations [33, 42, 43, 61], we can still clearly see differences between the simulated and experimental PDFs in Figure 2b-c and Figure S3 of the Supplementary Material. These discrepancies are attributed to a number of common limitations associated with force field MD simulations: (i) potential inaccuracy of the empirically derived force field parameters used in the MD simulations, (ii) the relatively small size of the simulation cell (i.e., ~30 × 30 × 30 Å$^3$) as compared with real samples, and (iii) the significantly faster cooling rates adopted in typical MD simulations (~ $10^{12}$ K/s) as compared with a typical experimental condition (1-100 K/s [46]). Another contributing factor is the presence of small crystalline impurities and trace elements (e.g., Fe, Ti and S) in the experimental samples that are not considered in the MD simulations [33].

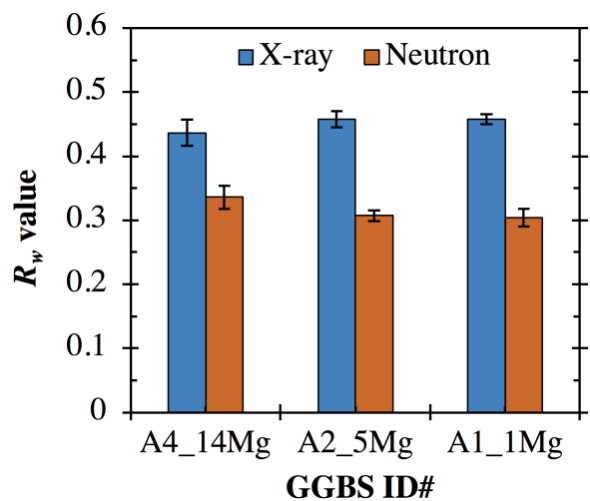

Figure 3. Agreement between experimental X-ray and neutron PDFs and simulated PDFs based on MD-generated structures for three CMAS compositions in Group A (Table 1), evaluated by the $R_w$ values (discussed in Section 4 of the Supplementary Material). The values reported in the figure are averages based on analysis of six MD-generated structural representations for each CMAS composition, with the error bars indicating one standard deviation.

Based on the MD trajectory of the last 500 ps of equilibration at 300K (500 structural snapshots), we calculated the partial radial distribution functions (RDFs) for the nearest atom-atom pairs (i.e., Si-O, Al-O, Mg-O, Ca-O and O-O), with the calculation details given in Section 4 of the Supplementary Material. The nearest interatomic distances for these atom-atom pairs are then determined from these partial RDFs (peak of each curve), as illustrated in Figure 4, where the typical partial RDFs for the atom-atom pairs (Si-O, Al-O, Mg-O, Ca-O and O-O) in a Group A glass (i.e., the A3_7Mg composition in Table 2) are given. The results for the nearest interatomic distances for all the CMAS glasses in Group A-C are summarized in Table 3, where it is clear that the moderate compositional variations of the CMAS glasses studied here have negligible impact on these nearest interatomic distances. However, we do observe obvious differences in the peak intensity of these partial RDF curves for the different glass compositions, as illustrated in Figure S4 of the Supplementary Material. These interatomic distances agree reasonably well with the corresponding experimental values for Si-O (~1.63 Å), Al-O (~1.75 Å), Mg-O (~2.00 Å), Ca-O (~2.35 Å) and O-O (~2.67 Å) in aluminosilicate glasses [33, 62, 63], with the differences smaller than ~3%. The largest deviation is seen for the Ca-O distance, where the MD-generated structures give an overestimation of ~0.07 Å. This overestimation of Ca-O distance is likely due to the Guillot force field [50] used here, where a similar overestimation has been previously reported in the literature for a comparable force field (e.g., Matsui [64]) [33].

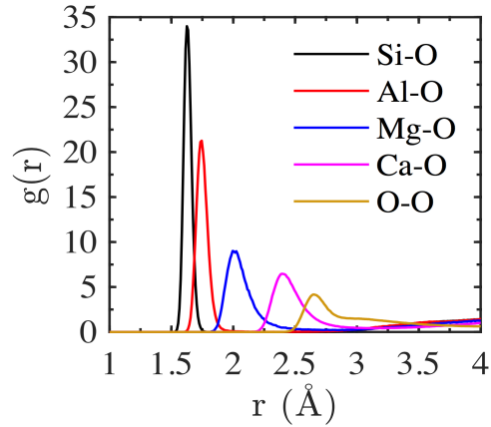

Figure 4. Partial radial distribution functions for the nearest atom-atom pairs of the A3_7Mg glass composition.

Table 3 also summarizes the average coordination number (CN) for the different atom-atom pairs using the cutoff distances of 2.2 Å, 2.5 Å, 2.9 Å and 3.2 Å for Si-O, Al-O, Mg-O and Ca-O correlations, respectively. These cutoff distances have been determined from the first minima in the partial RDFs and were kept the same for analysis of all the data (including the CAS glasses in the next section) for consistency and ease of comparison. The results show that the Si atoms in all the CMAS glasses investigated here are in IV-fold coordination, which is consistent with $^{29}$Si NMR data [40] and atomistic simulations [33, 45] on similar CMAS glasses. Al atoms are seen to be dominated by IV-fold coordination with a small percent of V-fold coordination (~0-7%, refer to the CN distributions for Al atoms in Figure S5a in the Supplementary Material). Based on classical glass theory [38], there should not be any V-fold Al in the CMAS glasses investigated here due to the large proportion of excess Ca and Mg cations beyond that required to charge-balance IV-fold alumina sites (i.e., $[Al(O_{1/2})_4]^{-1}$). However, many experimental and simulation findings [65-69] have challenged this classical view of the glass model by revealing the formation of a small proportion of high-coordination alumina sites (mainly V-fold) in peralkaline or peralkaline earth aluminosilicate glasses.

The Ca cations in the CMAS glasses are seen to have an average CN of ~6.7-6.8 (Table 3), and the CN distributions in Figure S5b of the Supplementary Material reveal the dominance of VI- and VII-fold coordination for all the CMAS glasses investigated here along with the presence of V-, VIII- and IX-fold CN, which is consistent with previous investigations on similar aluminosilicate glasses [33, 40, 45, 62, 70, 71]. The Mg cations have a smaller average CN (~4.9-5.2) than that of the Ca cations, with the CN distributions dominated by V-fold coordination and the simultaneous presence of IV- and VI-fold for all the CMAS glasses (Figure S5c of the Supplementary Material), which is also consistent with literature data on Mg coordination in Mg-containing silicate glasses [62, 70] (a brief summary has been given in ref. [33]). Overall, the results in Table 3 suggest that the MD-generated structural representations are able to capture the local atomic structure of CMAS glasses. Moreover, the compositional variation studied here has a moderate impact on the CN of Ca and Mg cations, and to a lesser extent the Al atom, yet as expected, its impact on the CN of Si atoms and the nearest interatomic distances is negligible.

Table 3. Nearest atom-atom interatomic distances and the coordination numbers (CNs) in the first coordination shell of Ca, Mg, Al and Si atoms for the different GGBS compositions in Groups A, B and C. The nearest atom-atom interatomic distances were obtained from the peak positions of the partial RDFs (as shown in Figure 4), while the CNs were calculated using cutoff distances of 2.2 Å, 2.5 Å, 2.9 Å and 3.2 Å for Si-O, Al-O, Mg-O and Ca-O pairs, respectively. The interatomic distance and CN values in the table are averages based on three separate partial RDFs (from the three MD trajectories), with one standard deviation given in the brackets (the values have been rounded to two decimal places).

| ID # | Calculated NBO/T! | Nearest interatomic distance (Å) | | | | | Average coordination number | | | |
| --- | --- | --- | --- | --- | --- | --- | --- | --- | --- | --- |
| | | Si-O | Al-O | Mg-O | Ca-O | O-O | Si | Al | Mg | Ca |
| A1_1Mg | 1.52 (0.02) | 1.63 (0.00) | 1.75 (0.00) | 2.04 (0.02) | 2.42 (0.00) | 2.69 (0.00) | 4.00 (0.00) | 4.03 (0.01) | 4.99 (0.07) | 6.75 (0.02) |
| A2_5Mg | 1.77 (0.01) | 1.63 (0.00) | 1.75 (0.00) | 2.03 (0.01) | 2.42 (0.00) | 2.69 (0.00) | 4.00 (0.00) | 4.02 (0.01) | 5.07 (0.10) | 6.76 (0.05) |
| A3_7Mg | 1.80 (0.03) | 1.63 (0.00) | 1.75 (0.00) | 2.03 (0.01) | 2.43 (0.00) | 2.68 (0.00) | 4.00 (0.00) | 4.03 (0.01) | 5.17 (0.05) | 6.75 (0.01) |
| A4_14Mg | 1.99 (0.03) | 1.63 (0.00) | 1.75 (0.00) | 2.04 (0.01) | 2.43 (0.00) | 2.68 (0.00) | 4.00 (0.00) | 4.04 (0.02) | 5.13 (0.02) | 6.80 (0.01) |
| B1_8Mg | 1.54 (0.00) | 1.63 (0.00) | 1.75 (0.00) | 2.03 (0.01) | 2.43 (0.00) | 2.68 (0.00) | 4.00 (0.00) | 4.04 (0.03) | 5.09 (0.06) | 6.74 (0.02) |
| B2_10Mg | 1.67 (0.01) | 1.63 (0.00) | 1.75 (0.00) | 2.04 (0.01) | 2.43 (0.01) | 2.69 (0.00) | 4.00 (0.00) | 4.03 (0.01) | 5.15 (0.09) | 6.78 (0.06) |
| B3_14Mg | 1.79 (0.02) | 1.63 (0.00) | 1.75 (0.00) | 2.04 (0.01) | 2.43 (0.00) | 2.69 (0.00) | 4.00 (0.00) | 4.03 (0.01) | 5.19 (0.05) | 6.83 (0.02) |
| C1_7Al | 1.25 (0.01) | 1.63 (0.00) | 1.75 (0.00) | 2.03 (0.01) | 2.43 (0.00) | 2.67 (0.00) | 4.00 (0.00) | 4.02 (0.01) | 5.04 (0.06) | 6.73 (0.01) |
| C2_14Al | 1.41 (0.02) | 1.63 (0.00) | 1.75 (0.00) | 2.04 (0.01) | 2.42 (0.00) | 2.69 (0.00) | 4.00 (0.00) | 4.02 (0.03) | 5.04 (0.09) | 6.72 (0.02) |
| C3_17Al | 1.84 (0.00) | 1.63 (0.00) | 1.75 (0.00) | 2.04 (0.01) | 2.42 (0.00) | 2.69 (0.00) | 4.00 (0.00) | 4.03 (0.00) | 5.14 (0.12) | 6.77 (0.03) |

! Average NBO/T value based on analysis of six structural representations from the MD simulations, with one standard deviation given in the bracket.

In contrast to the relatively small variation in the nearest interatomic distances and CNs for the CMAS glasses (Group A-C), the degree of depolymerization (NBO/T, calculated from the MD-generated structural representations) varies considerably depending on the composition, as also shown in Table 3. These calculated NBO/T values are compared with the theoretical NBO/T ratio estimated from simple stoichiometric arguments [38] in Figure 5, which shows that the calculated values are close to the theoretical estimations, with $R^2$ values of 0.99-1.00 for linear fits for each group (i.e., Group A, B and C). It is also seen that our simulations generally give slightly lower NBO/T values (up to ~10% difference) than the theory [38], which is consistent with our previous DFT calculations [33] as well as MD simulations in the literature on percalcic aluminosilicate glasses [71]. The likely reason for these lower NBO/T values is the formation of a small proportion of FO species not connected to any network formers (i.e., Si and Al) in our structural representations and the literature MD simulations, which are not accounted for in the classical glass theory (only considers NBO and bridging oxygen (BO) species bonded to two network formers in the first coordination shell) [38]. One possible formation reaction for FO species in highly percalcic aluminosilicate glasses, as suggested in ref. [72], is $2NBO \leftrightharpoons FO + BO$, which indicates that the formation of one FO consumes two NBOs. When taking into account the consumption of NBO species via the above reaction, the calculated (NBO+2FO)/T becomes exactly the same as the theoretical NBO/T, as illustrated in Figure 5. Note that all the values from the MD simulations in Figure 5 have very small standard deviations.

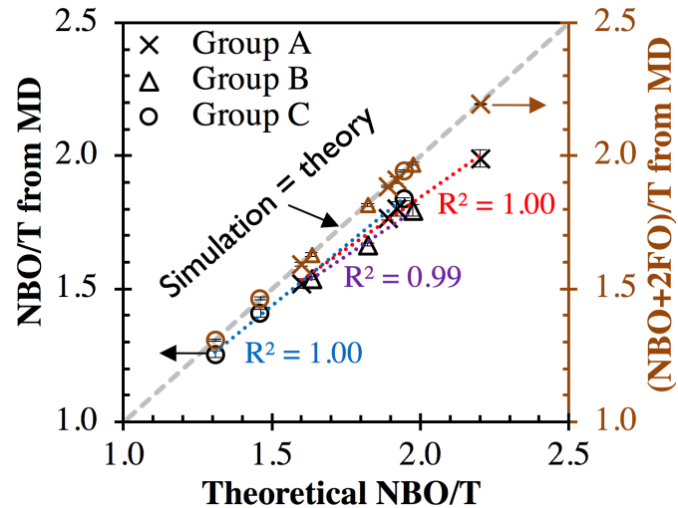

Figure 5. Comparison of the calculated NBO/T (left axis) and (NBO+2FO)/T (right axis) from MD-generated structural representations with the corresponding theoretical NBO/T values obtained from simple stoichiometric arguments [38]. NBO = non-bridging oxygen; FO = free oxygen. The error bars are one standard deviation based on the analysis of six structural representations. The $R^2$ values for the linear fits (red, purple and blue dotted lines for Group A, B and C, respectively) of the calculated NBO/T for each group of CMAS glasses are shown in the figure.

### 3.1.2 CAS glasses (Group D)

In contrast to Group A-C glasses which are representative of GGBS compositions with a relatively high degree of depolymerization (theoretical NBO/T =~ 1.3 to 2.2, Table 3) and low compositional variation, the CAS glasses in Group D cover a much wider compositional range. Specifically, D1-4 reside in the peraluminous region ($CaO/Al_2O_3 < 1$), where there are insufficient modifier cations (i.e., $Ca^{2+}$) to charge-balance the negatively charged alumina tetrahedra (i.e., $[Al(O_{1/2})_4]^{-1}$), and hence D1-4 are expected to be fully polymerized according to the stoichiometric argument (NBO/T = 0) [38]. D5-7 are slightly percalcic glasses (CaO/$Al_2O_3$ > 1) with theoretical NBO/T = ~ 0.1 to 0.32 and are expected to be slightly depolymerized. D8 is highly percalcic with a theoretical NBO/T of ~1.68, representing a highly depolymerized structure similar to the CMAS glasses in Group A-C. While D1-4 compositions are relevant to class F fly ash, D5-7 and D8 compositions are more representative of class C fly ash and GGBS compositions, respectively.

Table 4 summarizes the nearest interatomic distances and the average CN for the eight CAS glasses determined from the MD-generated structural representations. It is clear that the interatomic distances are similar to each other among the eight CAS glasses and are also similar to those of the CMAS glasses in Table 3. As expected, all the Si atoms are 100% IV-coordinated, while the Al CN is slightly higher than 4.0, indicative of a small proportion of Al in higher coordination. Compared with the CMAS glasses, the CAS glasses exhibit larger variation in the average CN of Al atoms (~4.02-4.11 in Table 4 as compared to ~4.02-4.04 in Table 3) owing to the larger extent of compositional variation. A larger variation in the Ca CN is also seen in Table 4 (~6.69-7.19) as compared to ~6.72-6.83 for the CMAS glasses in Table 3. The CN distribution for the Ca cations in the CAS glasses is seen to be dominated by VII-fold coordination with a considerable amount of VI- and VIII-fold except for the most peraluminous glass, i.e., D3, which is dominated by VI-fold coordination (see Figure S6 of the Supplementary Material). These results are similar to the CMAS glasses (Figure S5b of the Supplementary Material) and are generally consistent with literature data on calcium aluminosilicate glasses [73, 74].

It appears from Figure 6 that the average Al CN is, in general, inversely correlated with the theoretical NBO/T (negative value indicating insufficient modifier content for charge-balancing), with a lower theoretical NBO/T value generally leading to a higher average Al CN. This general trend is consistent with literature data [38] which show that the formation of high-coordination Al in CAS glasses increases in the highly peraluminous region (CaO/$Al_2O_3$ < 1). This is because there is a greater need for charging balancing in the highly peraluminous regions as there are insufficient charge-balancing cations (e.g., Ca), and the formation of high-coordination Al and tri-cluster oxygen are two postulated mechanisms for local charge-balancing in aluminosilicate glasses [75]. However, it is also seen in the intermediate region (theoretical NBO/T = − 0.1 to 0.32; D1, D4, D5,

D6 and D7) that there is an increasing trend of Al CN with increasing theoretical NBO/T value (the gray region in Figure 6), which seems to contradict the overall trend (black dashed line in Figure 6). A closer examination of the data in this intermediate region reveals an increasing trend of Al CN with an increasing amount of Ca cations, as highlighted by the light blue region in Figure 6. This deviation from the global trend in the intermediate region could be attributed to the increasing Ca content, since it has been shown in the literature that high strength modifier cations (e.g., Ca over Na) favor the formation of high-coordination Al [38].

Table 4. The nearest interatomic distances and the coordination numbers (CNs) in the first coordination shell of Ca, Al and Si atoms for the different CAS glass compositions in Group D. The nearest atom-atom interatomic distances were obtained from analysis of partial RDFs, while the coordination numbers were calculated using the same cutoff distances adopted for Group A-C. The interatomic distance and CN values in the table are averages based on three separate partial RDFs (from the three MD trajectories), with one standard deviation given in the brackets (the values have been rounded to two decimal places).

| ID # | Theoretical NBO/T[1] | Nearest interatomic distance (Å) | | | | Average coordination number | | |
|------|------|------|------|------|------|------|------|------|
| | | Si-O | Al-O | Ca-O | O-O | Si | Al | Ca |
| D1 | −0.10* | 1.63 (0.00) | 1.75 (0.00) | 2.45 (0.01) | 2.67 (0.00) | 4.00 (0.00) | 4.04 (0.005) | 6.69 (0.01) |
| D2 | −0.21* | 1.63 (0.00) | 1.75 (0.00) | 2.44 (0.01) | 2.68 (0.01) | 4.00 (0.00) | 4.11 (0.01) | 7.00 (0.09) |
| D3 | −0.32* | 1.63 (0.00) | 1.76 (0.00) | 2.44 (0.01) | 2.68 (0.00) | 4.00 (0.00) | 4.09 (0.01) | 6.49 (0.04) |
| D4 | −0.02* | 1.63 (0.00) | 1.75 (0.00) | 2.44 (0.01) | 2.68 (0.00) | 4.00 (0.00) | 4.06 (0.02) | 6.77 (0.08) |
| D5 | 0.32 | 1.63 (0.00) | 1.75 (0.00) | 2.43 (0.01) | 2.67 (0.00) | 4.00 (0.00) | 4.07 (0.03) | 7.00 (0.07) |
| D6 | 0.26 | 1.63 (0.00) | 1.74 (0.00) | 2.43 (0.01) | 2.69 (0.00) | 4.00 (0.00) | 4.07 (0.01) | 7.19 (0.03) |
| D7 | 0.10 | 1.63 (0.00) | 1.75 (0.00) | 2.43 (0.01) | 2.71 (0.00) | 4.00 (0.00) | 4.06 (0.01) | 7.11 (0.02) |
| D8 | 1.68 | 1.63 (0.00) | 1.74 (0.00) | 2.42 (0.00) | 2.69 (0.00) | 4.00 (0.00) | 4.02 (0.00) | 6.82 (0.02) |

[1] Theoretical NBO/T determined from chemical composition using simple stoichiometric argument [38], with the details given in Section 3 of the Supplementary Material.

* Peraluminous region with no NBO in theory. The negative values indicate that there are insufficient Ca cations to charge balance all the Al atoms, assuming all Al atoms are in IV-fold coordination. A more negative value indicates a greater Ca cation deficiency.

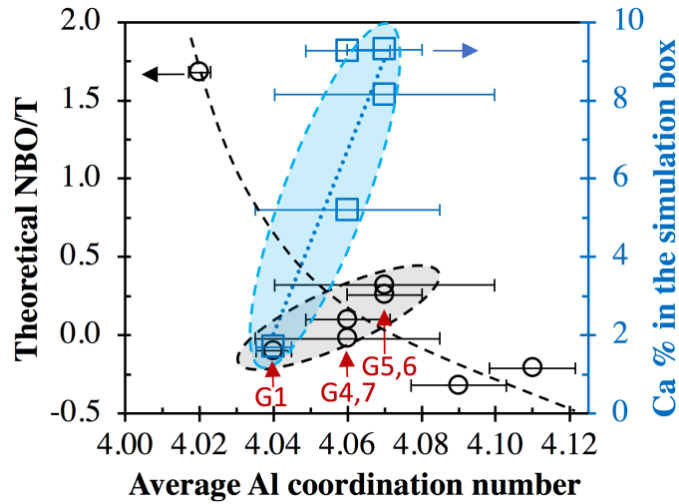

Figure 6. Correlation between theoretical NBO/T and the average Al CN calculated from MD simulations (black circle). The correlation between the average Al CN and Ca content (molar percentage) from the MD structural representations is also shown in the figure using blue squares (right axis). All the values are averages based on the analysis of six different structural representations, with one standard deviation shown in the figure. The dashed lines and shaded circles are given to guide the eye.

The CN distribution of Al atoms from the MD simulations is compared with the corresponding $^{27}$Al NMR data obtained from ref. [15] in Table 5. The simulation results are seen to agree reasonably well with the experimental data in the percalcic region (theoretical NBO/T > ~0.0, i.e., D5-8) as shown in Table 5 and Figure 7a. Also, both the experiment and simulation show that the CAS glass compositions in this region are dominated by IV-fold coordinated Al with less than ~10% V-fold and negligible VI-fold, which is consistent with other investigations on percalcic aluminosilicate glasses [38]. However, in the peraluminous region (theoretical NBO/T < ~0.0; D1-D4), the proportion of higher coordination Al (V- and VI-fold) is much lower in the simulation (~5-10%) than the corresponding experimental results (~9-33%), although the general Al coordination trend has been captured by the simulations as evidenced by the positive correlation between experiment and simulation for both IV- and V-fold coordination ($R^2$ values of 0.58-0.62 for linear fits, shown in Figure 7a). The discrepancies are especially large in the highly peraluminous region (e.g., D2 and D3), which can be partially attributed to the selected cutoff distance used during the calculation of CNs. As illustrated in Figure 7a, the proportion of IV- and V-fold Al atoms significantly decrease and increase, respectively, when a cutoff distance of 2.8 Å is used (as opposed to 2.5 Å). However, this increased cutoff distance also leads to an increase of V-fold and a decrease of IV-fold Al for the percalcic CAS glasses in Figure 7a. Hence, the root cause of this large discrepancy in the highly peraluminous region is likely the accuracy of force field adopted here for predicting Al coordination in this region, although several other limitations associated with MD simulations (as has been briefly outlined in Section 3.1.1) might have also contributed to the difference. Development of a force field that can accurately capture the Al coordination characteristics in both highly peraluminous and percalcic regions of CAS and CMAS glasses is outside the scope of the current investigation but is worth exploring in the future.

Table 5. Comparison of Al CN distribution with $^{27}$Al NMR results from ref. [15] for Group D glasses (CAS).

| ID # | This study (in %) | | | | NMR results from ref. [15] (in %) | | |
|------|------|------|------|------|------|------|------|
| | $^{III}$Al | $^{IV}$Al | $^{V}$Al | $^{VI}$Al | $^{IV}$Al | $^{V}$Al | $^{VI}$Al |
| D1 | 1.0 (0.4) | 94.2 (1.0) | 4.7 (0.7) | 0.0 (0.005) | 81 | 18 | 1 |
| D2 | 0.3 (0.3) | 89.5 (1.1) | 9.5 (1.1) | 0.7 (0.4) | 65 | 32 | 3 |
| D3 | 0.3 (0.2) | 90.6 (1.3) | 8.6 (1.0) | 0.5 (0.7) | 64 | 33 | 3 |
| D4 | 0.3 (0.2) | 94.0 (1.7) | 5.2 (1.7) | 0.5 (0.4) | 90 | 9 | 1 |
| D5 | 0.0 (0.0) | 93.5 (2.5) | 6.0 (2.0) | 0.5 (0.5) | 93 | 7 | 0 |
| D6 | 0.0 (0.0) | 93.5 (1.2) | 5.8 (1.7) | 0.7 (0.6) | 94 | 5 | 1 |
| D7 | 0.0 (0.0) | 93.9 (0.8) | 5.8 (0.6) | 0.3 (0.3) | 95 | 5 | 0 |
| D8 | 0.0 (0.0) | 98.0 (0.3) | 2.0 (0.3) | 0.0 (0.0) | 94 | 6 | 0 |

Comparison of the proportion of BO and NBO species from the MD simulations with those calculated from NMR data available in ref. [15] is given in Figure 7b. It is clear that the BO content from the simulations and experiments agrees reasonably well, with the differences smaller than ~8% for most of the glasses except for D8, where the difference is larger at ~24%. A strong linear correlation is also seen between the simulated and experimental BO content, with an $R^2$ value of 0.99, as shown in Figure 7b. Moreover, the simulated NBO content agrees reasonably well with the experimental NBO, except for the two highly peraluminous glasses (i.e., D2 and D3), where the simulations show that the quantities of NBO species are negligible (~0.5-1%), as would be expected for highly peraluminous glasses. In contrast, the experimental data indicate that a considerable amount of NBO species (~12-18%) has formed in these two glasses. In the CAS glass literature, ~3-6% of NBO species are often observed in tectosilicate compositions ($CaO/Al_2O_3 = 1$, and theoretical NBO/T of 0) with $^{17}O$ NMR measurements [38, 76], which is close to our MD simulation results (~5%) and the calculated NBO content from NMR data for D1 [15] (close to the tectosilicate composition with a theoretical NBO/T of ~ –0.02). However, $^{17}O$ NMR data on peraluminous CAS glasses [38] show that the proportion of NBO species decreases as the CAS glass becomes increasingly peraluminous and become undetectable (< 0.5%) at theoretical NBO/T values of ~ –0.18 to 0.24. This inconsistency between our simulation-derived NBO content and the experimental NBO content obtained from the modeling of $^{29}Si$ NMR spectra in the peraluminous region suggests possible inaccuracies associated with the fitting of the NMR data [15] given the overlapping spectra from different $Q$ species. In fact, we can clearly see differences between simulated and experimental $^{29}Si$ NMR spectra in ref. [15], especially for the highly peraluminous glasses (i.e., D2 and D3), which has been attributed to several simplified assumptions in the model (as discussed in ref. [15]). In addition to the clear discrepancies found in the NMR fitted spectra, limitations associated with MD simulations could also contribute to the differences seen between experimental and simulation results in Figure 7b. This includes the accuracy of the adopted force field and several other factors that have been briefly outlined in Section 3.1.1. In spite of these limitations, here the MD simulations have adequately captured the major structural features (i.e., the nearest interatomic distance, CNs and oxygen speciation) along with the anticipated composition-structure relationships for these CAS glasses.

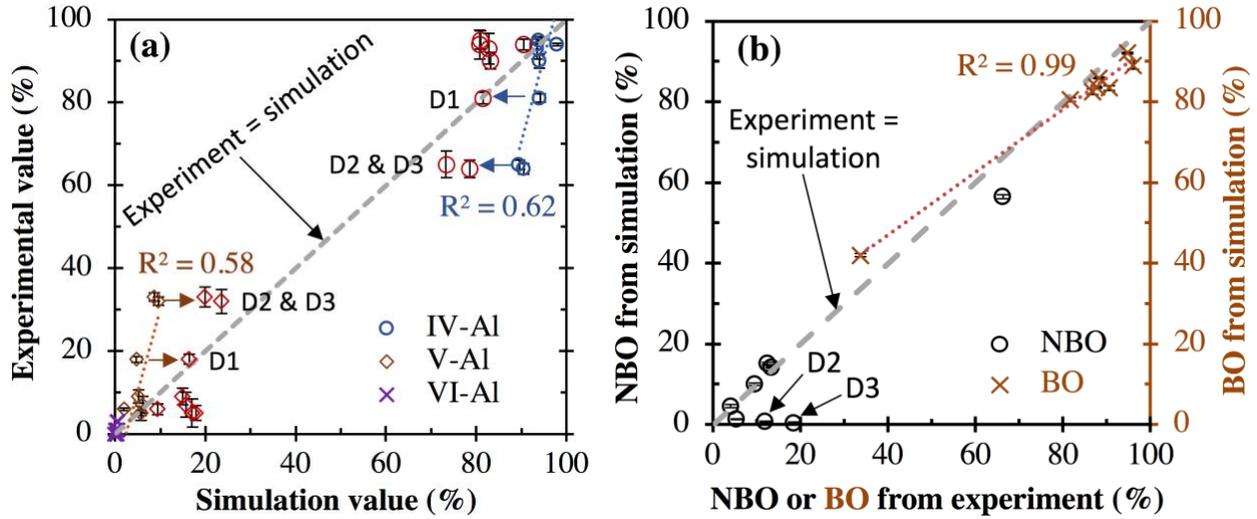

Figure 7. Comparison between experimental and simulation results for (a) different Al coordination and (b) BO and NBO content of the Group D CAS glasses (experimental results from ref. [15]). Linear fits of the IV- and V-fold Al contents are shown in (a) using dotted lines, with the goodness-of-fit $R^2$ values also given. The impact of increasing cutoff distance from 2.5 to 2.8 Å on the simulation-derived proportion of IV- and V-fold Al in the CAS glasses is also shown in (a) using red squares and circles, respectively. A linear fit of the BO content is given in (b) using a dotted line, with the $R^2$ value given in the figure. All the simulation results are averaged based on six structural representations, with one standard deviation given in the figure.

## 3.2 Structural descriptors for CMAS and CAS glass reactivity

The reactivity of CMAS and CAS glasses is important for their applications as SCMs in blended Portland cements and as precursor materials in AAM systems, yet the atomic origin controlling their reactivity is not well understood, as was briefly discussed in the Introduction. The chemical composition and atomic structure of these glassy phases have been seen to significantly impact their reactivity in the above applications, although other factors such as the particle size distribution, degree of amorphicity, solution chemistry and curing conditions can also have a profound impact on reactivity [1-3, 10, 36, 77]. In this investigation, we focus on understanding how the CMAS and CAS glass reactivity is influenced by the atomic structural attributes of the glass. Specifically, in the following section we evaluate how several structural descriptors derived from structural analysis (i.e., average metal oxide dissociation energy and degree of depolymerization) and dynamics analysis (i.e., average self-diffusion coefficient at melting) of the MD-generated structural representations from Section 3.1 correlate with different reactivity data obtained from the four literature investigations outlined in Table 1 (i.e., the corresponding CMAS and CAS glasses in Group A-D) [8, 11, 12, 15].

### 3.2.1 Average metal oxide dissociation energy (AMODE)

The dissolution of the CMAS and CAS glasses requires breaking of different metal-oxygen bonds (i.e., Si-O, Al-O, Mg-O and Ca-O) [32]. Given the oxygen CN number for each type of atom (i.e., Ca, Mg, Al and Si) from the MD simulations in Section 3.1 and single metal-oxygen bond strength from literature data, it is possible to derive a parameter that provides an overall estimate of the energy required to break/dissolve the oxide glass. This parameter, denoted as the average metal oxide dissociation energy (AMODE), is defined as follows (Equation 1):

$$AMODE = \frac{\sum N_M \cdot CN_M \cdot BS_{M-O}}{\sum N_M} \qquad (1)$$

where $N_M$ is the number of each type of metal cation ($M$ = Ca, Mg, Al, or Si) in the oxide glass, $CN_M$ and $BS_{M-O}$ are the average coordination number and the average metal-oxygen single bond strength (BS) for each type of atom $M$, respectively. The $CN_M$ values are calculated from the MD simulations (Tables 3 and 4), while the $BS_{M-O}$ values can be obtained from the literature. The BS of the single Si-O, Mg-O and Ca-O bonds in IV-, VI- and VI-fold coordination are ~106, ~37 and ~32 kilocalories, respectively [78]. The BS of the Al-O single bond depends highly on the Al CN: IV-fold Al has a BS of 79-101 kilocalories (the average value of 90 is taken here) while VI-fold Al has a BS of 53-67 kilocalories (the average of 60 is taken) [78]. The average BS of the single

Al-O bond is calculated by assuming that the BS of V-fold Al-O is the average of IV- and VI-fold Al-O (i.e., (90+60)/2 = 75 kilocalories). According to Figure S5 and S6 in the Supplementary Material, both Ca and Mg cations in the CMAS and CAS glasses have a distribution of CNs (with average values of ~6.7-7.5 and ~4.9-5.2, respectively, Tables 3 and 4), so the actual average BS for Ca-O and Mg-O bonds will be slightly different from those adopted here for VI-fold Ca and Mg (i.e., ~37 and ~32 kilocalories). However, their impact should be relatively small due to the significantly lower BS of Ca-O and Mg-O single bonds (as compared to Al-O and Si-O bonds).

Figure 8a-d shows how this AMODE parameter derived using Equation 1 correlates with the different reactivity data from refs. [8, 11, 12, 15] for Group A, B, C and D glasses, respectively. Although efforts were made in those experimental investigations to ensure similar particle sizes for each group, there is still ~2-20% difference in particle surface area within a group. A recent investigation on GGBS reactivity in alkaline environments showed that the reactivity (based on ICC measurements) increases linearly as a function of particle specific surface area ($R^2$ values of 0.97-1.00 for linear fits) [36]. Hence, all the experimental data presented in Figure 8 (and thereafter) have been normalized by the particle surface area within each group.

It is clear from Figure 8a that the AMODE of the four CMAS glasses in Group A is strongly and positively correlated with the time to reach the first reaction peak in the ICC data collected on $Na_2CO_3$-activated GGBSs (with an $R^2$ value of 0.95). A logarithmic scale of ICC time is used for the x-axis (as opposed to a linear scale adopted for other reactivity data in Figure 8b-d) because the extent of reaction (or ICC cumulative heat curve) is approximately a logarithmic function with time, as illustrated in Figure S7 of the Supplementary Material. Figure 8a shows that a ~3.3% increase in the AMODE value leads to a dramatic delay (over 30 hours) for the appearance of the first ICC peak. For the $Na_2CO_3$-activated GGBS system, the first ICC reaction peak is mainly associated with the formation of the initial reaction products (e.g., calcite and gaylussite) between the dissolved species from the neat GGBS (e.g., Ca species) and the carbonate species in the activator solution [8]. Hence, this suggests that the GGBS with a higher AMODE experiences significantly slower GGBS dissolution (e.g., release of Ca species) in these systems. This is consistent with our expectation since a higher AMODE value means that on average more energy is required to break/dissolve an oxide glass.

Figure 8b shows the correlation between the AMODE parameter for the CMAS glasses in Group B and the bound water content in the resulting $Na_2SiO_3$-activated GGBS obtained from thermogravimetric analysis (TGA), which is a reflection of the degree of reaction of GGBSs, defined as the percentage weight loss between 30 and 650 °C [12]. The bound water content data in Figure 8b (and thereafter) have been averaged over five data sets collected at different curing times to increase robustness, and this does not change the overall trend seen among individual data set as illustrated in Figure S8 of the Supplementary Material. It is clear from Figure 8b that the bound water content in the $Na_2SiO_3$-activated GGBS is strongly and inversely correlated with the AMODE value of the CMAS glassy phase in the GGBS (with an $R^2$ value of 0.95 for a linear fit). A decrease in AMODE is seen to lead to a higher degree of reaction and hence a higher reactivity, which is consistent with the results in Figure 8a. A similar trend is seen for alkali-activated GGBSs based on the CMAS glass compositions in Group C (Figure 8c, with an $R^2$ value of 0.93).

Figure 8d shows the relationship between the AMODE parameter of the eight synthetic CAS glasses in Group D and the extent of reaction of these glasses in a blended mixture of NaOH, $Ca(OH)_2$ and $CaCO_3$ (reacted for 180 days), obtained from quantitative XRD analysis [15]. Due to the larger compositional range in Group D, these glasses exhibit a wider range of AMODE values (i.e., ~320-400) than the CMAS glasses in Group A-C (i.e., ~300-315). Despite the wider compositional range of Group D glasses, the AMODE parameter is seen to be almost linearly and inversely correlated with the extent of reaction data from quantitative XRD analysis, possessing an $R^2$ value of 0.97 for a linear fit (Figure 8d). The CAS glass with a lower AMODE value is seen to exhibit a substantially higher degree of reaction after 180 days and hence a higher reactivity in the blended Portland alkaline environment. This trend is also consistent with the results for the CMAS glasses in Group A-C (Figure 8a-c).

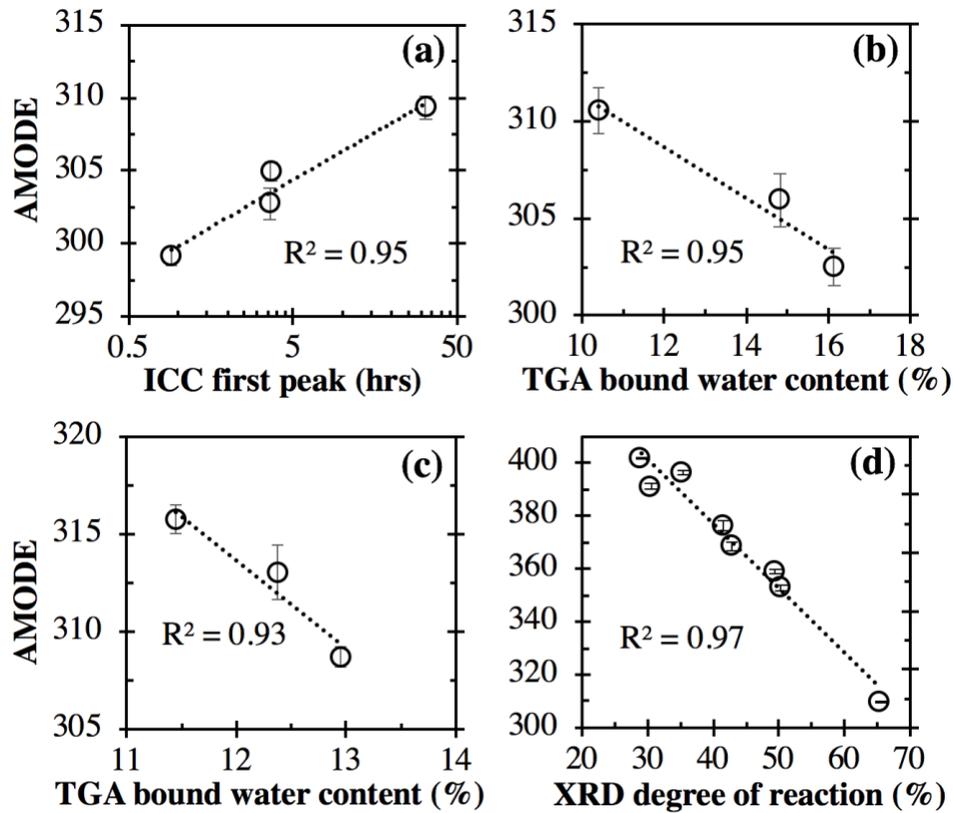

Figure 8. Correlation between the average metal oxide dissociation energy (AMODE) parameter of the CMAS and CAS glasses and the reactivity data collected for the corresponding aluminosilicate glasses [8, 11, 12, 15] for (a) Group A, (b) Group B, (c) Group C and (d) Group D. The isothermal conduction calorimetry (ICC) data (time to reach the first ICC peak) were obtained from ref. [8] based on $Na_2CO_3$-activated GGBS with the same chemical composition as the CMAS glasses in Group A. The thermogravimetric analysis (TGA) bound water content data in (b) and (c) were obtained from refs. [12] and [11] on $Na_2SiO_3$-activated GGBS with Group B and C chemical compositions, respectively. The extent of reaction data in (d) were obtained from ref. [15] based on quantitative X-ray diffraction (XRD) analysis of synthetic CAS glasses in Group D activated by a blended mixture of NaOH, $Ca(OH)_2$ and $CaCO_3$. A linear fit between the AMODE parameter and the reactivity data (dotted line) is given in each figure (note that the x-axis for (a) is logarithmic), with the $R^2$ value (goodness of fit) also given. The error bars are one standard deviation based on the analysis of six structural representations from three independent MD production runs.

Overall, the results in Figure 8 show that the AMODE parameter gives an accurate description of the relative reactivity of the CMAS and CAS glasses when exposed to alkaline environments. We have also used this AMODE parameter to correlate with other reactivity data (specifically the extent of reaction from NMR and/or thermodynamic modeling, compressive strength data, and TGA bound water data collected for NaOH-activated samples) available in refs. [8, 11, 12, 15]. The results are presented in Figure S9 of the Supplementary Material, and the level of agreement as evaluated by the $R^2$ values are generally comparable with those presented in Figure 8 for each group of glass. The performance of the AMODE parameter is encouraging, particularly given the inherent uncertainty of the experimental measurements and data analysis process (e.g., XRD phase quantification), along with several limitations associated with the calculations of the AMODE parameter: (i) the potential deviation of the actual average BS of the single Mg-O and Ca-O bonds from those adopted here for VI-fold Mg and Ca cations, (ii) the approximation made with the BS of Al-O in Al polyhedra (in particular, V-fold Al), and (iii) the potential inaccuracies of the estimated CNs from MD simulations especially for Al atoms in the highly peraluminous region as discussed in Section 3.1.2. The ability for AMODE to predict relative reactivity for the synthetic CAS glasses in Group D is especially noteworthy since this group span a much wider compositional range and does not exhibit obvious compositional inter-correlation as seen for the CMAS glasses in Group A-C (see Figure 1 and Figures S1 and S2 of the Supplementary Material). Furthermore, AMODE is seen to perform much better than the NBO/T for describing CAS glass reactivity (NBO/T shown in ref. [15]), where the NBO/T is determined by considering V- and VI-fold Al (quantified from $^{27}$Al NMR data) as network modifiers (more details have been given in ref. [15]).

### 3.2.2    Self-diffusion coefficient at melting

At temperatures above the melting point of the CMAS and CAS glasses, the mobility of atoms increases dramatically due to ongoing making and breaking of metal-oxygen bonds in the melt, in a sense similar to the metal-oxygen bond-breaking process during glass dissolution. With this in mind, we have calculated the mean square displacement (MSD) of the different elements in each glass at 2000 K as a function of time using the MD trajectories from the *NVT* equilibration step at 2000 K. The MSD results for a typical CMAS glass melt are shown in Figure 9, which clearly reveal that the two modifier cations (i.e., Ca and Mg) exhibit much higher mobility (larger MSD values at a given time) than the network formers (i.e., Si and Al atoms). This is expected as the Al-O and Si-O bonds are much stronger (hence harder to break) than the Ca-O and Mg-O bonds (as shown in Section 3.2.1). We also see that the MSD (i.e., mobility) of the Al is noticeably higher than that of Si, which is also attributed to the lower average BS of the Al-O bond compared with Si-O bond, as discussed in Section 3.2.1. In contrast, the Mg cation is seen to exhibit a slightly higher mobility than the Ca cation although the BS of the Mg-O bond (~37 kcal) is slightly higher than that of Ca-O in VI-fold coordination (~32 kcal). This could be attributed to the higher average coordination of Ca cation that requires breaking of more Ca-O bonds for a Ca cation to move around, as compared to the case of a Mg cation. Furthermore, the smaller size of $Mg^{2+}$ (~0.80 Å for V-fold [79]) compared with $Ca^{2+}$ (~1.14-1.20 Å for VI- and VII-fold [79]) may contribute to the higher mobility of the Mg cation since a smaller size makes diffusion easier.

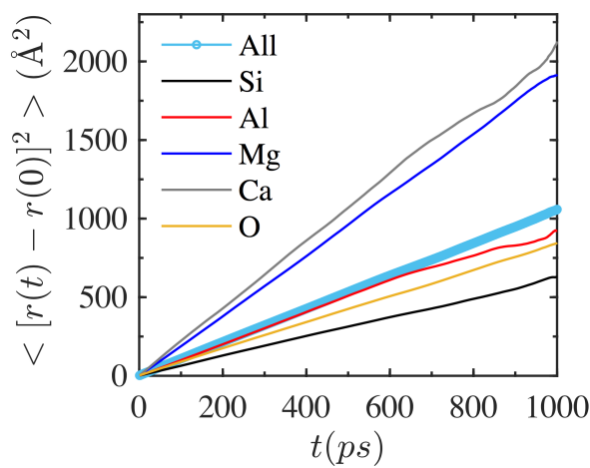

Figure 9. Mean square displacement (MSD) of each element along with the average of all atoms denoted as "All" in a typical CMAS glass (i.e., A3_7Mg) as a function of time during the 1 ns of MD equilibration step at 2000 K.

Based on the MSD results, we can calculate the average self-diffusion coefficient (ASDC) for all the atoms in each glass using Einstein's equation (Equation 2):

$$D = \frac{\langle [r(t) - r(0)]^2 \rangle}{6t} \tag{2}$$

where D is the self-diffusion coefficient, $t$ is the simulation time, and $\langle [r(t) - r(0)]^2 \rangle$ is the MSD between time $t$ and 0. Hence, $D$ is related to the slope of the MSD curve. To improve accuracy, we have chosen the most linear portion of the data for all the calculations (i.e., 50-700ps).

The ASDC parameter is a measure of the average atomic mobility of all the atoms in each glass simulated at 2000 K and hence, in a sense, reflects the ease of bond-breaking in the glass (i.e., higher mobility ≈ easier to break bonds and dissolve glass). Although there are some similarities between the ASDC and AMODE parameters, the main difference is that calculation of ASDC does not involve any assumptions with BS for the different metal-oxygen bonds whereas the AMODE parameter is a more direct measurement of the ease of bond-breaking. Figure 10 illustrates how the ASDC parameter correlates with the different reactivity data for the CMAS and CAS glasses investigated here (same experimental data as reported in Section 3.2.1, obtained from refs. [8, 11, 12, 15]). It is clear that a high degree of correlation is achieved using this ASDC parameter for the different reactivity data, with $R^2$ values of 0.92-0.99. The trends in Figure 10 are opposite to those shown in Figure 8, with a higher ASDC value exhibiting a higher reactivity, as expected. The similar $R^2$ values suggest a similar level of predictive performance for both parameters. In fact, we see from Figure S10 in the Supplementary Material that the ASDC parameter is almost linearly correlated with the AMODE parameter for all the glass compositions studied here, with an $R^2$ value of 0.99 using a linear fit.

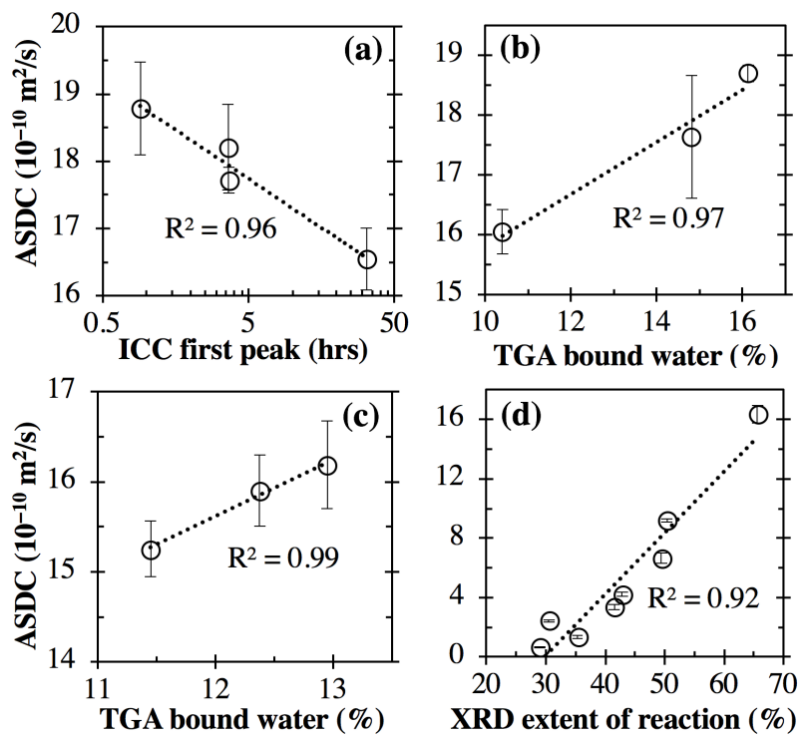

Figure 10. Correlation between the average self-diffusion coefficient (ASDC) of all the atoms in each CMAS and CAS glass at a temperature of 2000 K and reactivity data for the corresponding aluminosilicate glasses from (a) Group A, (b) Group B, (c) Group C and (d) Group D. Details about these experimental data [8, 11, 12, 15] have been given in the caption of Figure 8. The error bars are one standard deviation based on the analysis of six structural representations from three independent MD production runs.

We have also examined the degree of correlation of this ASDC parameter with other experimental data collected on the CMAS and CAS glasses in refs. [8, 11, 12, 15], including the extent of reaction from NMR and/or thermodynamic modeling results, compressive strength data, and TGA bound water data collected for NaOH-activated samples. The results are shown in Figure S11 of the Supplementary Material, where a linear correlation similar to that in Figure 10 is observed for most of the data. These additional analyses in the Supplementary Material (including Figures S9 and S11) reinforce the observations that both the AMODE and ASDC parameters give an accurate indication of relative reactivity for the CMAS and CAS glasses studied here.

Finally, the performance of the commonly used degree of polymerization (i.e., NBO/T) and the modified (NBO+2FO)/T (both calculated from MD simulations results) has also been evaluated with respect to the experimental data for Group A-D, with the findings presented in Figures S12-14 of the Supplementary Material. The level of correlation for the four data sets in Figure S12 achieved using NBO/T and (NBO+2FO)/T are compared with those of the AMODE and ASDC parameters in Table 6. For the CMAS glasses in Group A-C, the $R^2$ values achieved with NBO/T and (NBO+2FO)/T (0.83-0.99) are comparable with, or only slightly lower than, those obtained using the AMODE (0.93-0.95) and ASDC (0.96-0.99) parameters. However, the $R^2$ value achieved with NBO/T (0.74) for the CAS glasses in Group D is much lower than the other two parameters (0.97 and 0.92 for AMODE and ASDC, respectively). The generally poorer performance of NBO/T and (NBO+2FO)/T is attributed to the fact that these two parameters does not distinguish between the type of network former (i.e., IV-fold Al versus Si atoms) and the type of network modifier (i.e., Ca versus Mg cations) with respect to reactivity. In contrast, these potential differences between former/modifier type have been explicitly (and implicitly) accounted for by the AMODE (and ASDC) parameter introduced in the previous sections. The observation that the NBO/T and (NBO+2FO)/T parameters perform better for the CMAS glasses in Group A-C (as compared to Group D) may be partially attributed to the fact that the oxide compositions within each group of CMAS glass are highly correlated with each other as shown in Figure 1 and Figure S1 of the Supplementary Material. A more detailed discussion of the performance of NBO/T and (NBO+2FO)/T parameters is given in Section 12 of the Supplementary Material.

Table 6. Summary of the level of agreement ($R^2$ values) achieved for a linear regression between the NBO/T or (NBO+2FO)/T parameter and the four experimental data sets associated with Group A-D (see Figure S12 in the Supplementary Material for linear fits), in comparison with the AMODE and ASDC parameters and their associated level of agreement (see Figures 8 and 10, respectively).

| Parameter | $R^2$ value for linear regression | | | |
| --- | --- | --- | --- | --- |
| | Group A | Group B | Group C | Group D |
| NBO/T | 0.99 | 0.91 | 0.85 | 0.74 |
| (NBO+2FO)/T | 0.98 | 0.94 | 0.83 | 0.73 |
| AMODE | 0.95 | 0.95 | 0.93 | 0.97 |
| ASDC | 0.96 | 0.97 | 0.99 | 0.92 |

### 3.3 Broader impact & limitations

#### 3.3.1 Broader impact

The development of accurate structural descriptors that are able to connect CAS and CMAS glass compositions with reactivity (and other properties) is critical to a number of important industrial applications, including blended Portland cements and AAMs. In this investigation, we developed two structural descriptors based on force field MD simulations, which exhibit superior performance for describing a range of reactivity data collected for a variety of CMAS and CAS compositions (as compared to the commonly used degree of depolymerization parameter). On one hand, this method can be readily extended to cover more complex aluminosilicate glasses, including those containing $Fe_2O_3$, $K_2O$, $Na_2O$, $MnO$ and $TiO_2$. This would allow the impact of all the oxide components to be explicitly or implicitly incorporated into these two structural descriptors for a more holistic description of the reactivity behavior of these highly complex glassy materials. On the other hand, this method may also be extended to describe the chemical durability and reactivity of other types of glasses or minerals, including those utilized in nuclear waste encapsulation, bioglass dissolution and carbon mineralization. Nevertheless, the limitations associated with application of this method needs to be carefully considered, as outlined in the next section.

#### 3.3.2 Limitations

Several limitations associated with this investigation warrant discussion. First, the reactivity of amorphous aluminosilicates in alkaline environments is highly complex and a number of other factors (in addition to the composition and structure discussed in this investigation) could have a large impact on reactivity, such as activator solution chemistry, particle size distribution, degree of amorphicity, curing conditions, and phase segregation in the original glass [1-3, 10, 36, 77]. The complexity further increases for many SCMs used in blended cements and precursor materials used for AAM synthesis (e.g., coal-derived fly ash), which are often more heterogeneous and complex in composition and mineralogy than those presented in this investigation (which are either pure synthetic glasses or GGBSs with a high level of amorphicity). The potential phase segregation in the glassy phases of the SCMs (or precursor materials), as has been shown to be the case for fly ash [17, 80], could have a dramatic impact on the reactivity of SCMs (or precursor materials) in alkaline environments. Hence, the above factors need to be taken into account when applying the structural descriptors proposed in this investigation.

Second, the accuracy of force field MD simulations is highly dependent on the force field used, hence a large portion of this investigation focused on evaluating the performance of the Guillot force field [50] and specifically its ability to capture the structural features of CAS and CMAS glasses seen in experiments. As seen in the Section 3.1, although this force field can capture many of the structural features, it is not able to give an accurate prediction of Al coordination in highly peraluminous regions (albeit the general trend is captured). This necessitates future development or optimization of force field parameters to provide a more accurate description of Al coordination in both percalcic and peraluminous regions. In addition, MD simulations bear several common limitations that have been briefly discussed in this investigation, including fast cooling rate and limited cell size (as compared to real samples). Finally, as already discussed in Section 3.2.1, the calculation of the AMODE parameter relies on several assumptions on the bond strength of individual metal-oxygen bonds, especially the Al-O bonds for IV-, V- and VI-fold Al. More accurate prediction of the bond strength of the different metal-oxygen bonds for a range of coordination states would be helpful for future investigations.

## 4    Conclusions

The composition-structure-property relationships for amorphous aluminosilicates in alkaline environments are important for many industrial applications, including blended cements and alkali-activated materials. In this investigation, we employed force field-based molecular dynamics (MD) simulations to generate detailed structural representations for CaO-MgO-SiO$_2$-Al$_2$O$_3$ (CMAS) and CaO-SiO$_2$-Al$_2$O$_3$ (CAS) glasses with compositions similar to ten GGBSs and eight synthetic glasses reported in the literature. We showed that the glass structural representations obtained using the MD simulations agree reasonably well with our experimental X-ray and neutron pair distribution function data of select CMAS compositions, as well as literature data, in terms of the nearest interatomic distance, coordination number, and degree of depolymerization. Based on the structural analysis results, we developed two new structural descriptors and evaluated their ability to predict relative reactivity for the CMAS/CAS glass compositions, specifically (i) the average metal oxide dissociation energy (AMODE), an estimate of the average energy required to break/dissolve all the metal-oxygen bonds in the glass, and (ii) the average self-diffusion coefficient (ASDC) for all the atoms in the glass melt at 2000 K, which is a reflection of the overall atomic mobility and hence easiness to break metal-oxygen bonds.

Connecting these structural descriptors with different reactivity data from four literature investigations, including isothermal conduction calorimetry (ICC), bound water content from thermogravimetric analysis (TGA), and the extent of reaction from quantitative X-ray diffraction analysis, shows that the two parameters exhibit strong correlations with almost all the experimental data for the CMAS glass compositions studied here with $R^2$ values close to or higher than 0.90. For the CAS glasses which span a wider compositional range than the CMAS glasses considered here, the AMODE and ASDC parameters exhibit much stronger correlations with the corresponding reactivity data than the degree of depolymerization (NBO/T) parameter. This behavior is attributed to the fact that the AMODE (and ASDC) parameter has explicitly (and implicitly) taken into account the differences in the ease of breaking the various metal-oxygen bonds in the glasses, which is not considered by the commonly used NBO/T parameter. The results strongly suggest that the AMODE and ASDC parameters are promising structural descriptors that connect CMAS and CAS glass compositions with their reactivity in alkaline environments, and, therefore, this investigation serves as a crucial step forward in establishing the important composition-structure-reactivity relationships for amorphous aluminosilicates in alkaline environments, relevant to AAMs and blended Portland cements.

# 5    Supplementary Material

Correlation between the different oxide content of the GGBSs and synthetic CAS glasses; Estimation of the CMAS glass density at different temperatures; Theoretical estimation of NBO/T; Calculation of PDFs, partial RDFs, and $R_w$; Comparison of simulated and experimental PDFs; Comparison of the partial RDFs; CN distributions for the CMAS and CAS glasses; Evolution of select reactivity data as a function of reaction time; Correlation between the AMODE parameter and the additional experimental data; Correlation between the AMODE parameter and the average self-diffusion coefficient (ASDC) at melting; Correlation between the ASDC at melting and the additional experimental data; Performance of the degree of depolymerization.

# 6    Acknowledgments

This material is based on work supported by ARPA-E under Grant No. DE-AR0001145 and the National Science Foundation under Grant No. 1362039. K.G. was partially supported by a Charlotte Elizabeth Proctor Fellowship from the Princeton Graduate School. The MD simulations were performed on computational resources supported by the Princeton Institute for Computational Science and Engineering (PICSciE) and the Office of Information Technology's High Performance Computing Center and Visualization Laboratory at Princeton University. The 11-ID-B beam line is located at the Advanced Photon Source, an Office of Science User Facility operated for the U.S. DOE Office of Science by Argonne National Laboratory, under U.S. DOE Contract No. DE-AC02-06CH11357. The NPDF instrument is located at Los Alamos Neutron Science Center, previously funded by DOE Office of Basic Energy Sciences. Los Alamos National Laboratory is operated by Los Alamos National Security LLC under DOE Contract DE-AC52-06NA25396. The upgrade of NPDF was funded by the NSF through grant DMR 00-76488.

# 7    References

1.    J.L. Provis and J.S.J. van Deventer, eds. *Alkali activated materials: state-of-the-art report, RILEM TC 224-AAM*. 2014, Springer/RILEM: Dordrecht.

2.  J.L. Provis and S.A. Bernal, Geopolymers and related alkali-activated materials, Annu. Rev. Mater. Res., 2014. **44**: p. 299-327.

3.  J. Skibsted and R. Snellings, Reactivity of supplementary cementitious materials (SCMs) in cement blends, Cem. Concr. Res., 2019. **124**(105799): p. 1-16.

4.  R. Snellings, G. Mertens, and J. Elsen, Supplementary cementitious materials, Rev. Mineral. Geochem., 2012. **74**(1): p. 211-278.

5.  P.J.M. Monteiro, S.A. Miller, and A. Horvath, Towards sustainable concrete, Nat. Mater., 2017. **16**(7): p. 698-699.

6.  K.L. Scrivener, V.M. John, and E.M. Gartner, Eco-efficient cements: Potential economically viable solutions for a low-$CO_2$ cement-based materials industry, Cem. Concr. Res., 2018. **114**: p. 2-26.

7.  K. Gong and C.E. White, Impact of chemical variability of ground granulated blast-furnace slag on the phase formation in alkali-activated slag pastes, Cem. Concr. Res., 2016. **89**: p. 310-319

8.  X. Ke, S.A. Bernal, and J.L. Provis, Controlling the reaction kinetics of sodium carbonate-activated slag cements using calcined layered double hydroxides, Cem. Concr. Res., 2016. **81**: p. 24-37.

9.  S.A. Bernal, R. San Nicolas, R.J. Myers, R.M. de Gutiérrez, et al., MgO content of slag controls phase evolution and structural changes induced by accelerated carbonation in alkali-activated binders, Cem. Concr. Res., 2014. **57**: p. 33-43.

10. M. Ben Haha, G. Le Saout, F. Winnefeld, and B. Lothenbach, Influence of activator type on hydration kinetics, hydrate assemblage and microstructural development of alkali activated blast-furnace slags, Cem. Concr. Res., 2011. **41**(3): p. 301-310.

11. M. Ben Haha, B. Lothenbach, G. Le Saout, and F. Winnefeld, Influence of slag chemistry on the hydration of alkali-activated blast-furnace slag — Part II: Effect of $Al_2O_3$, Cem. Concr. Res., 2012. **42**(1): p. 74-83.

12. M. Ben Haha, B. Lothenbach, G. Le Saout, and F. Winnefeld, Influence of slag chemistry on the hydration of alkali-activated blast-furnace slag — Part I: Effect of MgO, Cem. Concr. Res., 2011. **41**(9): p. 955-963.

13. P.Z. Wang, R. Trettin, and V. Rudert, Effect of fineness and particle size distribution of granulated blast-furnace slag on the hydraulic reactivity in cement systems, Adv. Cem. Res., 2005. **17**(4): p. 161-167.

14. B.O. Mysen and P. Richet, *Silicate glasses and melts*. 2nd ed. 2018, Amsterdam, Netherlands: Elsevier.

15. S. Kucharczyk, M. Zajac, C. Stabler, R.M. Thomsen, et al., Structure and reactivity of synthetic CaO-$Al_2O_3$-$SiO_2$ glasses, Cem. Concr. Res., 2019. **120**: p. 77-91.

16. R. Snellings, Surface chemistry of calcium aluminosilicate glasses, J. Am. Ceram. Soc., 2015. **98**(1): p. 303-314.

17. P.T. Durdziński, R. Snellings, C.F. Dunant, M. Ben Haha, et al., Fly ash as an assemblage of model Ca–Mg–Na-aluminosilicate glasses, Cem. Concr. Res., 2015. **78**: p. 263-272.

18. Y. Feng, Q. Yang, Q. Chen, J. Kero, et al., Characterization and evaluation of the pozzolanic activity of granulated copper slag modified with CaO, J. Clean. Prod., 2019. **232**: p. 1112-1120.

19. J.S.J. van Deventer, C.E. White, and R.J. Myers, A Roadmap for Production of Cement and Concrete with Low-$CO_2$ Emissions, Waste Biomass Valori., 2020: p. 1-31.

20. F. Bonk, J. Schneider, M.A. Cincotto, and H. Panepucci, Characterization by multinuclear high-resolution NMR of hydration products in activated blast-furnace slag pastes, J. Am. Ceram. Soc., 2003. **86**(10): p. 1712-1719.

21. A. Fernandez-Jimenez, F. Puertas, I. Sobrados, and J. Sanz, Structure of calcium silicate hydrates formed in alkaline-activated slag: influence of the type of alkaline activator, J. Am. Ceram. Soc., 2003. **86**(8): p. 1389-1394.

22. F. Puertas, M. Palacios, H. Manzano, J.S. Dolado, et al., A model for the C-A-S-H gel formed in alkali-activated slag cements, J. Eur. Ceram. Soc., 2011. **31**(12): p. 2043-2056.

23. R.J. Myers, S.A. Bernal, J.D. Gehman, J.S.J. van Deventer, et al., The role of Al in cross-linking of alkali-activated slag cements, J. Am. Ceram. Soc., 2015. **98**(3): p. 996-1004.

24. K. Yang and C.E. White, Multiscale pore structure determination of cement paste via simulation and experiment: The case of alkali-activated metakaolin, Cem. Concr. Res., 2020. **137**: p. 106212.

25. A. Blyth, C.A. Eiben, G.W. Scherer, and C.E. White, Impact of activator chemistry on permeability of alkali-activated slags, J. Am. Ceram. Soc., 2017. **100**: p. 1-12.

26. J. Osio-Norgaard, J.P. Gevaudan, and W.V. Srubar III, A review of chloride transport in alkali-activated cement paste, mortar, and concrete, Constr. Build. Mater., 2018. **186**: p. 191-206.

27. J.L. Provis, R.J. Myers, C.E. White, V. Rose, et al., X-ray microtomography shows pore structure and tortuosity in alkali-activated binders, Cem. Concr. Res., 2012. **42**(6): p. 855-864.

28.   J. Zhang, C. Shi, Z. Zhang, and Z. Ou, Durability of alkali-activated materials in aggressive environments: A review on recent studies, Constr. Build. Mater., 2017. **152**: p. 598-613.

29.   S.A. Bernal and J.L. Provis, Durability of alkali-activated materials: Progress and perspectives, J. Am. Ceram. Soc., 2014. **97**(4): p. 997-1008.

30.   E. Douglas and J. Brandstetr, A preliminary study on the alkali activation of ground granulated blast-furnace slag, Cem. Concr. Res., 1990. **20**(5): p. 746-756.

31.   S.L. Brantley, *Kinetics of mineral dissolution*, in *Kinetics of water-rock interaction*. 2008, Springer. p. 151-210.

32.   E.H. Oelkers, General kinetic description of multioxide silicate mineral and glass dissolution, Geochim. Cosmochim. Ac., 2001. **65**(21): p. 3703-3719.

33.   K. Gong, O.V. Özçelik, K. Yang, and C.E. White, Density functional modeling and total scattering analysis of the atomic structure of a quaternary CaO-MgO-Al$_2$O$_3$-SiO$_2$ (CMAS) glass: Uncovering the local environment of calcium and magnesium, Phys. Rev. Mater., 2021. **5**(1).

34.   C. Jiang, K. Li, J. Zhang, Q. Qin, et al., The effect of CaO(MgO) on the structure and properties of aluminosilicate system by molecular dynamics simulation, J. Mol. Liq., 2018. **268**: p. 762-769.

35.   C. Shi, D. Roy, and P. Krivenko, *Alkali-activated cements and concretes*. 2006, Abingdon, UK: Taylor & Francis.

36.   S. Blotevogel, A. Ehrenberg, L. Steger, L. Doussang, et al., Ability of the R3 test to evaluate differences in early age reactivity of 16 industrial ground granulated blast furnace slags (GGBS), Cem. Concr. Res., 2020. **130**: p. 105998.

37.   K. Aughenbaugh, T. Williamson, and M. Juenger, Critical evaluation of strength prediction methods for alkali-activated fly ash, Mater. Struct., 2015. **48**(3): p. 607-620.

38.   L.M. Thompson and J.F. Stebbins, Non-bridging oxygen and high-coordinated aluminum in metaluminous and peraluminous calcium and potassium aluminosilicate glasses: High-resolution [17]O and [27]Al MAS NMR results, Am. Mineral., 2011. **96**(5-6): p. 841-853.

39.   I. Pignatelli, A. Kumar, M. Bauchy, and G. Sant, Topological control on silicates' dissolution kinetics, Langmuir, 2016. **32**(18): p. 4434-4439.

40.   K. Shimoda, Y. Tobu, K. Kanehashi, T. Nemoto, et al., Total understanding of the local structures of an amorphous slag: perspective from multi-nuclear ([29]Si, [27]Al, [17]O, [25]Mg, and [43]Ca) solid-state NMR, J. Non-Cryst. Solids, 2008. **354**(10): p. 1036-1043.

41.     D.R. Neuville, L. Cormier, and D. Massiot, Al coordination and speciation in calcium aluminosilicate glasses: Effects of composition determined by $^{27}$Al MQ-MAS NMR and Raman spectroscopy, Chem. Geol., 2006. **229**(1): p. 173-185.

42.     C.E. White, J.L. Provis, T. Proffen, D.P. Riley, et al., Combining density functional theory (DFT) and pair distribution function (PDF) analysis to solve the structure of metastable materials: the case of metakaolin, Phys. Chem. Chem. Phys., 2010. **12**(13): p. 3239-3245.

43.     C. Siakati, R. Macchieraldo, B. Kirchner, F. Tielens, et al., Unraveling the nano-structure of a glassy CaO-FeO-SiO$_2$ slag by molecular dynamics simulations, J. Non-Cryst. Solids, 2020. **528**: p. 119771.

44.     A. Peys, C.E. White, D. Olds, H. Rahier, et al., Molecular structure of CaO–FeOx–SiO2 glassy slags and resultant inorganic polymer binders, J. Am. Ceram. Soc., 2018. **101**(12): p. 5846-5857.

45.     K. Shimoda and K. Saito, Detailed structure elucidation of the blast furnace slag by molecular dynamics simulation, ISIJ Int., 2007. **47**(9): p. 1275-1279.

46.     X. Li, W. Song, K. Yang, N.M.A. Krishnan, et al., Cooling rate effects in sodium silicate glasses: Bridging the gap between molecular dynamics simulations and experiments, J. Chem. Phys., 2017. **147**(7): p. 074501.

47.     N. Jakse, M. Bouhadja, J. Kozaily, J.W.E. Drewitt, et al., Interplay between non-bridging oxygen, triclusters, and fivefold Al coordination in low silica content calcium aluminosilicate melts, Appl. Phys. Lett., 2012. **101**(20): p. 201903.

48.     K. Baral, A. Li, and W.-Y. Ching, Ab initio molecular dynamics simulation of Na-doped aluminosilicate glasses and glass-water interaction, AIP Adv., 2019. **9**(7): p. 075218.

49.     A. Pedone and M.C. Menziani, *Computational modeling of silicate glasses: a quantitative structure-property relationship perspective*, in *Molecular Dynamics Simulations of Disordered Materials*. 2015, Springer. p. 113-135.

50.     B. Guillot and N. Sator, A computer simulation study of natural silicate melts. Part I: Low pressure properties, Geochim. Cosmochim. Ac., 2007. **71**(5): p. 1249-1265.

51.     QuantumATK version P-2019.03. *QuantumWise A/S*. Synopsys QuantumATK (https://www.synopsys.com/silicon/quantumatk.html) 2019  [cited 2020.

52.     J. Schneider, J. Hamaekers, S.T. Chill, S. Smidstrup, et al., ATK-ForceField: a new generation molecular dynamics software package, Model. Simul. Mater. Sc., 2017. **25**(8): p. 085007.

53.     P.J. Chupas, K.W. Chapman, and P.L. Lee, Applications of an amorphous silicon-based area detector for high-resolution, high-sensitivity and fast time-resolved pair distribution function measurements, J. Appl. Crystallogr., 2007. **40**(3): p. 463-470.

54.     T. Proffen, T. Egami, S.J.L. Billinge, A.K. Cheetham, et al., Building a high resolution total scattering powder diffractometer–upgrade of NPD at MLNSC, Appl. Phys. A-Mater., 2002. **74**(1): p. s163-s165.

55.     K. Gong and C.E. White, Nanoscale chemical degradation mechanisms of sulfate attack in alkali-activated slag, J. Phys. Chem. C, 2018. **122**(11): p. 5992–6004.

56.     T. Egami and S.J.L. Billinge, *Underneath the Bragg peaks: Structural analysis of complex materials*. 2003, Elmsford, NY: Pergamon.

57.     P. Juhás, T. Davis, C.L. Farrow, and S.J.L. Billinge, PDFgetX3: a rapid and highly automatable program for processing powder diffraction data into total scattering pair distribution functions, J. Appl. Cryst., 2013. **46**(2): p. 560-566.

58.     C.L. Farrow, P. Juhas, J.W. Liu, D. Bryndin, et al., PDFfit2 and PDFgui: Computer programs for studying nanostructure in crystals, J. Phys.: Condens. Matter, 2007. **19**(33): p. 335219.

59.     P.F. Peterson, M. Gutmann, T. Proffen, and S.J.L. Billinge, PDFgetN: a user-friendly program to extract the total scattering structure factor and the pair distribution function from neutron powder diffraction data, J. Appl. Crystallogr., 2000. **33**(4): p. 1192-1192.

60.     K. Page, C.E. White, E.G. Estell, R.B. Neder, et al., Treatment of hydrogen background in bulk and nanocrystalline neutron total scattering experiments, J. Appl. Crystallogr., 2011. **44**(3): p. 532-539.

61.     C.E. White, N.J. Henson, L.L. Daemen, M. Hartl, et al., Uncovering the true atomic structure of disordered materials: The structure of a hydrated amorphous magnesium carbonate ($MgCO_3 \cdot 3D_2O$), Chem. Mater., 2014. **26**(8): p. 2693-2702.

62.     M. Guignard and L. Cormier, Environments of Mg and Al in $MgO–Al_2O_3–SiO_2$ glasses: A study coupling neutron and X-ray diffraction and reverse Monte Carlo modeling, Chem. Geol., 2008. **256**(3-4): p. 111-118.

63.     L. Cormier, D.R. Neuville, and G. Calas, Structure and properties of low-silica calcium aluminosilicate glasses, J. Non-Cryst. Solids, 2000. **274**(1-3): p. 110-114.

64.     M. Matsui, A transferable interatomic potential model for crystals and melts in the system $CaO-MgO-Al_2O_3-SiO_2$, Mineral. Mag., 1994. **58**: p. 571-572.

65.     D.R. Neuville, L. Cormier, V. Montouillout, P. Florian, et al., Amorphous materials: Properties, structure, and durability: Structure of Mg-and Mg/Ca aluminosilicate glasses:

$^{27}$Al NMR and Raman spectroscopy investigations, Am. Mineral., 2008. **93**(11-12): p. 1721-1731.

66.     S.K. Lee, G.D. Cody, and B.O. Mysen, Structure and the extent of disorder in quaternary (Ca-Mg and Ca-Na) aluminosilicate glasses and melts, Am. Mineral., 2005. **90**(8-9): p. 1393-1401.

67.     M.J. Toplis, S.C. Kohn, M.E. Smith, and I.J.F. Poplett, Fivefold-coordinated aluminum in tectosilicate glasses observed by triple quantum MAS NMR, Am. Mineral., 2000. **85**(10): p. 1556-1560.

68.     S.K. Lee, H.-I. Kim, E.J. Kim, K.Y. Mun, et al., Extent of disorder in magnesium aluminosilicate glasses: insights from $^{27}$Al and $^{17}$O NMR, J. Phys. Chem. C, 2015. **120**(1): p. 737-749.

69.     M. Ren, J.Y. Cheng, S.P. Jaccani, S. Kapoor, et al., Composition–structure–property relationships in alkali aluminosilicate glasses: A combined experimental–computational approach towards designing functional glasses, J. Non-Cryst. Solids, 2019. **505**: p. 144-153.

70.     N. Trcera, D. Cabaret, S. Rossano, F. Farges, et al., Experimental and theoretical study of the structural environment of magnesium in minerals and silicate glasses using X-ray absorption near-edge structure, Phys. Chem. Miner., 2009. **36**(5): p. 241-257.

71.     L. Mongalo, A.S. Lopis, and G.A. Venter, Molecular dynamics simulations of the structural properties and electrical conductivities of CaO-MgO-Al$_2$O$_3$-SiO$_2$ melts, J. Non-Cryst. Solids, 2016. **452**: p. 194-202.

72.     H.W. Nesbitt, G.S. Henderson, G.M. Bancroft, R. Sawyer, et al., Bridging oxygen speciation and free oxygen (O$^{2-}$) in K-silicate glasses: Implications for spectroscopic studies and glass structure, Chem. Geol., 2017. **461**: p. 13-22.

73.     D.R. Neuville, L. Cormier, A.-M. Flank, V. Briois, et al., Al speciation and Ca environment in calcium aluminosilicate glasses and crystals by Al and Ca K-edge X-ray absorption spectroscopy, Chem. Geol., 2004. **213**(1-3): p. 153-163.

74.     L. Cormier and D.R. Neuville, Ca and Na environments in Na$_2$O–CaO–Al$_2$O$_3$–SiO$_2$ glasses: influence of cation mixing and cation-network interactions, Chem. Geol., 2004. **213**(1-3): p. 103-113.

75.     T. Charpentier, K. Okhotnikov, A.N. Novikov, L. Hennet, et al., Structure of strontium aluminosilicate glasses from molecular dynamics simulation, neutron diffraction, and nuclear magnetic resonance studies, J. Phys. Chem. B, 2018. **122**(41): p. 9567-9583.

76.     J.F. Stebbins and Z. Xu, NMR evidence for excess non-bridging oxygen in an aluminosilicate glass, Nature, 1997. **390**(6655): p. 60.

77.    K. Gong, Y. Cheng, L.L. Daemen, and C.E. White, *In situ* quasi-elastic neutron scattering study on the water dynamics and reaction mechanisms in alkali-activated slags, Phys. Chem. Chem. Phys., 2019(21): p. 10277-10292.

78.    K.H. Sun, Fundamental condition of glass formation, J. Am. Ceram. Soc., 1947. **30**(9): p. 277-281.

79.    R.D. Shannon, Revised effective ionic radii and systematic studies of interatomic distances in halides and chalcogenides, Acta Crystall. A-Crys., 1976. **32**(5): p. 751-767.

80.    P.T. Durdziński, C.F. Dunant, M. Ben Haha, and K.L. Scrivener, A new quantification method based on SEM-EDS to assess fly ash composition and study the reaction of its individual components in hydrating cement paste, Cem. Concr. Res., 2015. **73**: p. 111-122.